\documentclass[preprint,12pt,authoryear]{elsarticle}

\pdfoutput=1

\journal{Information and Software Technology}

\usepackage{balance}
\usepackage[utf8]{inputenc}
\usepackage{hyperref}
\usepackage{amssymb}
\usepackage{float}
\usepackage{multicol}
\usepackage[normalem]{ulem}
\usepackage[switch]{lineno}
\usepackage{tikz}
\usetikzlibrary{mindmap,shadows}
\definecolor{light-gray}{rgb}{0.56, 0.74, 0.56}

\biboptions{authoryear}

\newcommand{\FR}{\ensuremath{\mathit{FR}}}
\newcommand{\Savg}{\ensuremath{\overline{S}}}
\newcommand{\Smed}{\ensuremath{\widetilde{S}}}
\newcommand{\STaLiRo}{S-TaLiRo\xspace}
\newcommand{\ARIsTEO}{ARIsTEO\xspace}
\newcommand{\staliro}{\STaLiRo}

\newcommand{\FalStar}{FalStar\xspace}
\newcommand{\foresee}{ForeSee\xspace}

\newcommand{\falsify}{falsify\xspace}

\newcommand{\FalCAuN}{FalCAuN\xspace}

\usepackage{amsmath,amsfonts}
\usepackage{esvect}
\usepackage{graphicx}
\usepackage{subfigure}
\usepackage{booktabs}
\usepackage{tabularx}
\usepackage{multirow}
\usepackage{balance}
\usepackage{xcolor}
\usepackage{cleveref}
\usepackage{enumitem}

\usepackage{mathtools}
\usepackage{wrapfig}
\usepackage{ragged2e}
\usepackage[all]{nowidow}
\usepackage{stfloats}
\usepackage{anyfontsize}
\usepackage{algorithm}
\usepackage{tikz}
\usepackage{amsmath}
\usepackage{color}
\usepackage{colortbl}
\usepackage{xspace}
\usetikzlibrary{positioning}
\usepackage{graphicx}

\begin{document}

\newboolean{showcomments}
\setboolean{showcomments}{true} % toggle to show or hide comments
\ifthenelse{\boolean{showcomments}}
{\newcommand{\nb}[3]{
  \fcolorbox{black}{#3}{\bfseries\sffamily\scriptsize#1}
  {\sf\small$\blacktriangleright$\textit{#2}$\blacktriangleleft$}
 }
 \newcommand{\version}{\emph{\scriptsize$-$working$-$}}
}
{\newcommand{\nb}[3]{}
 \newcommand{\version}{}
}

%  \rwcomment{text}{identified}{commentref}
\newcommand\addcitation[3]{\linelabel{#2} \textcolor{blue}{\cite{#1}} \comm{C\ref{#3}} \linelabel{#2end}  }
\newcommand\deletecitation[3]{\linelabel{#2} \textcolor{blue}{\sout{\mbox{\cite{#1}}}} \comm{C\ref{#3}} \linelabel{#2end}  }
\newcounter{commentnumber}
\newcommand\delete[3]{\linelabel{#2}\textcolor{blue}{\sout{#1}} \comm{C\ref{#3}} \linelabel{#2end}  }
\newcommand\rep[4]{\linelabel{#3}\textcolor{blue}{#1} \textcolor{blue}{\sout{#2}} \comm{C\ref{#4}} \linelabel{#3end}  }
\newcommand\change[3]{\linelabel{#2}\textcolor{blue}{#1} \comm{C\ref{#3}} \linelabel{#2end}  }

\newcommand\textreference[3]{\linelabel{#2} #1 \comm{C\ref{#3}} \linelabel{#2end}  }

\newcommand{\comm}[1]{\fcolorbox{black}{yellow}{\bfseries\sffamily\scriptsize#1}}

\begin{frontmatter}



\title{Reflections on Surrogate-Assisted Search-Based Testing: A Taxonomy and Two Replication Studies based on Industrial ADAS and Simulink Models}


\author[1]{Shiva Nejati\corref{cor1}}
\ead{snejati@uottawa.ca}

\author[2]{Lev Sorokin}
\ead{sorokin@fortiss.org}

\author[2]{Damir Safin}
\ead{safin@fortiss.org}

\author[3]{Federico Formica}
\ead{formicaf@mcmaster.ca}

\author[3]{Mohammad~Mahdi~Mahboob}
\ead{mahbom2@mcmaster.ca}

\author[3,4]{Claudio Menghi}
\ead{claudio.menghi@unibg.it}

\affiliation[1]{organization={University of Ottawa},
             addressline={Canada},
             city={Ottawa},
             postcode={K1N6N5},
             country={Canada}}

% \affiliation[3]{organization={McMaster University},
%             addressline={1280 Main St W},
%             city={Hamilton},
%             postcode={L8S4L8},
%            country={Canada}}
\affiliation[2]{organization={Fortiss},
             city = {Munich},
             country={Germany}}
            
 \affiliation[3]{organization={McMaster University},
 addressline={1280 Main St W},
   city ={Hamilton},
    postcode={ON L8S4L8},
country={Canada}
}

 \affiliation[4]{organization={ University of Bergamo},
             addressline={via Salvecchio 19},
             city={Bergamo},
             postcode={24129},
            country={Italy}}
            
\cortext[cor1]{Corresponding author}

\begin{abstract}
\emph{Context.} Surrogate-assisted search-based testing (SA-SBT) aims to reduce the computational time for testing compute-intensive systems. Surrogates enhance testing techniques by improving test case generation focusing the testing budget on the most critical portions of the input domain. In addition, they can serve as approximations of the system under test (SUT) to predict test results instead of executing the tests on compute-intensive SUTs.

\emph{Objective.} This article reflects on the existing SA-SBT techniques, particularly those applied to system-level testing and often facilitated using simulators or complex test beds. Recognizing the diversity of heuristic algorithms and evaluation methods employed in existing SA-SBT techniques, our objective is to synthesize these differences and present a comprehensive view of SA-SBT solutions. In addition, by critically reviewing our previous work on SA-SBT, we aim to identify the limitations in our proposed  algorithms and evaluation methods and to propose potential improvements.

\emph{Method.}  We present a taxonomy that categorizes and contrasts existing SA-SBT solutions and highlights key research gaps. To identify the evaluation challenges,  we conduct two replication studies of our past SA-SBT solutions: One study uses industrial advanced driver assistance system (ADAS)  and the other relies on a Simulink model benchmark. We compare our results with those of the original studies and identify the difficulties in evaluating SA-SBT techniques, including the impact of different contextual factors on results generalization and the validity of our evaluation metrics.

\emph{Results.} Based on our taxonomy and replication studies, we propose future research directions, including re-considerations in the current evaluation metrics used for SA-SBT solutions, utilizing surrogates for fault localization and repair in addition to testing, and creating frameworks for large-scale experiments by applying SA-SBT to multiple SUTs and simulators.

\end{abstract}



\end{frontmatter}


\section{Introduction}
\label{sec:intro}

To test complex and emerging systems at scale, the use of simulators is necessary. Simulators provide a flexible, effective, and efficient infrastructure to ensure the safety of complex systems compared to testing deployed systems in their operational environment (e.g., vehicles' on-road testing). However, executing each simulation scenario still takes a non-negligible amount of time, particularly when simulators need to be real-time such as in the case of simulators for autonomous vehicles. Since simulators' input spaces are very large and high-dimensional, we require techniques to make simulation-based testing scalable. For example, non-trivial simulations of an industrial model of a satellite system, capturing the satellite behavior for 24h, take, on average, around 84 minutes ($\sim$1.5 hours)~\cite{9283957}. To test this satellite model for all of its parameters and inputs, we need to execute hundreds of such simulations, which can take months or even years to complete. 

Search-based testing (SBT) has been traditionally used as an effective and efficient guidance for test generation performed at system-level~\cite{DBLP:conf/icse/Zeller17}. SBT algorithms provide flexibility in representing the input search space and  choosing a strategy to traverse a subset of the search space. To scale SBT to handle compute-intensive  cyber-physical systems (CPS) with large and multi-dimensional input spaces, a promising approach is to combine SBT with \emph{surrogate models}~\cite{jin2011surrogate} that enable effectively guiding the search exploration.

Surrogate models can enhance SBT in two ways: (1)~They identify the most critical regions of the input domain (i.e., the regions that include tests revealing  critical faults) and  sample test inputs that are in these regions. (2)~They predict simulation results instead of computing them. In both cases, surrogate models are generated and iteratively refined  based on the intermediary test executions. As surrogate models improve through successive refinements, their predictions can help improve the search guidance, as described by the first use case above. Alternatively, their predictions are used in lieu of actual test executions, as suggested by the second use case above, to reduce the search computation time by  executing fewer tests.

This article reflects on four papers published between 2014 and 2021~\cite{Matinnejad:14,Abdessalem:16,Abdessalem:18,9283957} that propose surrogate-assisted search-based testing (SA-SBT) techniques for  autonomous systems. The papers evaluated their proposed testing techniques on industrial 
Advanced Driving Assistance Systems (ADAS)~\cite{Abdessalem:16,Abdessalem:18}, a satellite system, and an open-source benchmark of Simulink models~\cite{9283957} as well as an industrial controller from the automotive domain~\cite{Matinnejad:14}.
Since these papers are the main subjects of our analysis, 
we refer to these four papers as the \emph{subject papers} hereafter in this article.

The subject papers apply SBT enhanced using surrogate models to system-level testing and demonstrate that this combination is effective for complex, emerging case studies such as self-driving and satellite systems. 
These subject papers complement existing SBT techniques that focus on testing software code and unit-level testing, and  the research on testing and verification of CPS based on formal methods~\cite{Clarke:01}. 
 These four papers, while studying CPS with different behaviours and varying degrees of autonomy and complexity, motivate the need for SA-SBT techniques as follows: In the context of CPS and given the safety-critical nature of these systems, system-level  testing is typically performed using simulators. CPS simulators, particularly those that are real-time, are compute-intensive, and their input spaces are large and high-dimensional. SA-SBT techniques are then proposed to scale simulation-based testing for CPS. 
 
 %Some summary of what we want to do in this reflection study: 
 The four subject papers adopt an empirical approach and evaluate their SA-SBT techniques using case studies from different domains. The techniques that these papers propose, although all can be categorized as SA-SBT, vary in several details. For example, three of the subject papers~\cite{Matinnejad:14,Abdessalem:16,9283957} use surrogates to reduce the computation time of computing fitness functions, while the fourth paper~\cite{Abdessalem:18} uses surrogates to guide the search exploration more effectively.  Given the differences in case studies and proposed techniques, it is uncertain how the results can be generalized to similar systems.  In this reflection paper, we review our previous research and attempt to answer two questions: (1)~What are the different types and variations of SA-SBT techniques? and (2)~To what extent can we reproduce the results of these four subject papers? To answer the first question, we propose a taxonomy for SA-SBT techniques and use it to categorize the subject papers and some recent papers that build on our research. To answer the second question, we report two replication studies that attempt to reproduce the results of two of the subject papers~\cite{Abdessalem:18,9283957}. In particular, one replication study applies the SA-SBT solution proposed by Abdessalem et. al.~\cite{Abdessalem:18} to an industrial ADAS. For the second replication study, we present the results obtained by  \ARIsTEO~\cite{9283957}, an approximation-refinement testing technique based on surrogate models, in three consecutive editions (2020~\cite{DBLP:conf/arch/ErnstABDFFMMPPY20}, 2021~\cite{DBLP:conf/arch/ErnstABCDFFG0KM21}, and 2022~\cite{ARCH22:ARCH_COMP_2022_Category_Report}) of the ARCH competition~\cite{ARCHWEBSITE}.
 The ARCH competition is a friendly yearly competition between testing tools for continuous and hybrid systems~\cite{ARCHWEBSITE}.

%To scale simulation-based testing, we proposed to combine SBT with Machine Learning supervised techniques, known as \emph{surrogate models}~\cite{jin2011surrogate}.

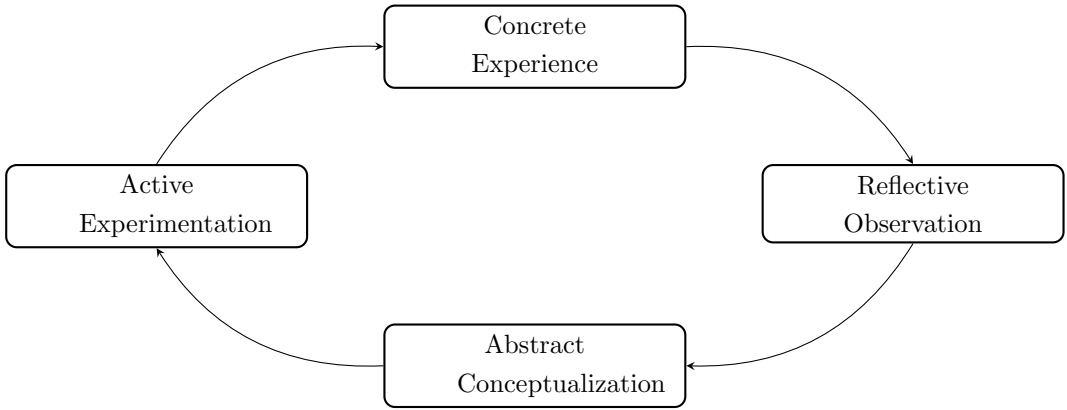
\begin{figure}[t]
    \centering
\begin{tikzpicture}[auto,
 block/.style ={rectangle, draw=black, thick, fill=white!20, text width=5em,align=center, rounded corners},
 block1/.style ={rectangle, draw=blue, thick, fill=blue!20, text width=5em,align=center, rounded corners, minimum height=2em},
 line/.style ={draw, thick, -latex',shorten >=2pt},
 cloud/.style ={draw=red, thick, ellipse,fill=red!20,
 minimum height=1em}]

\node [block, minimum width=4cm] (Experimentation) {\footnotesize Active \\ Experimentation\phantom{xxxxxx}};
\node [above right=1cm and 1cm of Experimentation, block, minimum width=4cm] (Experience) {\footnotesize Concrete  \\ Experience};
\node [block,below right=1cm and 1cm of Experimentation,minimum width=4cm] (Conceptualization) {\footnotesize Abstract\\ Conceptualization};
\node [block,below right =1cm and 1cm 
 of Experience,minimum width=4cm] (Reflective) {\footnotesize Reflective\\ Observation};

\path[-stealth,bend left] (Experimentation.north) edge (Experience.west);
\path[-stealth,bend left] (Experience.east) edge (Reflective.north);
\path[-stealth,bend left] (Reflective.south) edge (Conceptualization.east) ;
\path[-stealth,bend left] (Conceptualization.west) edge (Experimentation.south) ;
 \end{tikzpicture}
%    \includegraphics[width=0.9\columnwidth]{figures/kolb.pdf}
    %\imagecaptionspace
    \caption{Experiential learning cycle.}\label{fig:kolb}
\end{figure}

We use Kolb's model of experiential learning~\cite{DBLP:journals/software/DybaMG14} as a framework for our reflection study, which includes a taxonomy and two replication experiments. The Kolb model, shown in Figure~\ref{fig:kolb}, outlines the steps of experiential learning and emphasizes the role of reflection in turning direct experiences into abstract concepts and knowledge that drive further research. The process starts with obtaining concrete experience, followed by reflective observations in light of existing knowledge. Reflection should focus on any discrepancies between existing knowledge and the new experience. From reflections, the researcher develops observations into abstract concepts and generalizations, and tests these through active experimentation.

%To motivate and  structure our reflection study which is centered around a taxonomy and two replication experiments, we adopt the Kolb's model of experiential learning~\cite{DBLP:journals/software/DybaMG14}.  Kolb's model, shown in Figure~\ref{fig:kolb}, summarizes the  elements of experiential learning, and in particular, highlights the role of reflection in converting observations from direct experiences into abstract concepts and knowledge that can fuel further research.   Experiential learning  can help improve or further existing knowledge from direct experiences.  The process in Figure~\ref{fig:kolb} starts with obtaining \emph{concrete experience},  followed by the researchers' \emph{reflective observations} on the new experience and in  light of their existing knowledge. The reflection should particularly focus on any inconsistencies between the existing knowledge and new experience. Through reflection, the researcher develops the observations obtained from the experience into a collection of \emph{abstract concepts} (analysis) and generalizations. The last step is about testing newly created or modified concepts and hypotheses  through \emph{active experimentation}.

We structure our reflection study following the experiential learning loop in Figure~\ref{fig:kolb}.  We develop a taxonomy 
(Section~\ref{sec:taxonomy})  that  provides a set of abstract concepts and categories to help researchers classify and compare different SA-SBT research. We use the taxonomy to reflect on the current state of research and practice in the SA-SBT field, and to  identify current research gaps (Section~\ref{sec:position}). We replicate two of our subject papers~\cite{Abdessalem:18,9283957} and present our observations based on the replication studies (Section~\ref{sec:replications}). Our findings, which include the taxonomy, the categorization of the state-of-art along with the taxonomy dimensions, and two replication studies lead us to identify a number of shortcomings and challenges that impact the research on SA-SBT. We discuss future research directions that the software testing researchers can embark on to improve and unify  the research in this field. Section~\ref{sec:lessons} presents our recommendations for future research which are listed as follows: (1)~standardizing metrics for evaluating SA-SBT solutions, (2)~using surrogates to improve fault localization and repair, (3)~creating frameworks for large-scale and consistent application of SA-SBT across multiple SUTs and simulators, and (4)~developing benchmarks and organizing international competitions to promote the continuous replication of research tools.

\section{Taxonomy}
\label{sec:taxonomy}
\begin{figure}[H]
\begin{tikzpicture}[ every annotation/.style = {draw,
                     fill = black, font = \scriptsize}]
  \path[mindmap,concept color=light-gray,text=black,
    every node/.style={concept,circular drop shadow,font=\scriptsize},
    root/.style    = {concept color=light-gray,
      font=\scriptsize,text width=1em,minimum size=4em},
    level 1 concept/.append style={font=\scriptsize,
      sibling angle=45,text width=2em,
    level distance=7em,inner sep=0pt,minimum size=4em},
    level 2 concept/.append style={font=\scriptsize,level distance=7em, sibling angle=37,minimum size=2.2em},
  ]
  node[root] {SA-SBT} 
  [clockwise from=90]
    child[concept color=light-gray] {
      node[concept] {Purpose}
        [clockwise from=130]
      child { node[concept] (LaTeXForum)
        {Fitness Computation}}
      child { node[concept] (LaTeXArtikel)
        {Test\\ Generation\\ Guidance} }
      child { node[concept] (LaTeXNews)
        {Individuals\\ Selection}}
    }
    child[concept color=light-gray] {
      node {Usage} [clockwise from=65]
        child { node (goForum) {Test case generation -- single} }
        child { node (goWiki) {Test case filtering} }
        child { node (goWiki) {Test case generation -- ensemble}}
    }
    child[concept color=light-gray] {
      node {Subject} [clockwise from=2]
        child { node (goForum) {Model} }
        child { node (goWiki) {Model\&\\
        Req.-Ass} }
    }
    child[concept color=light-gray] {
      node[concept] {Design}
        [clockwise from=-23]
      child { node[concept] (TeXnique)
        {Manual} }
      child { node[concept] (TeXweltQA)
        {Automatic} 
              node[concept] {Automatic}
        [clockwise from=-10]
      child { node[concept] (TeXnique)
        {Static} }
         child { node[concept] (TeXnique)
        {Dynamic} }
        }
    }
    child[concept color=light-gray]{
    node[concept] {Type}
        [clockwise from=-50]
    child[concept color=light-gray] {
      node[concept] {Model Type}
        [clockwise from=330]
      child { node[concept] (TikZGalerie) 
        {White Box} }
      child { node[concept] (TeXampleBlog)
        {Grey Box} }
    child { node[concept] (TeXampleBlog)
        {Black Box} }
    }
    child[concept color=light-gray] {
      node[concept] (PGFPlots) {Output Type}
      [clockwise from=270]
      child[concept color=light-gray] { node[concept] (TeXweltQA)
        {Single Data} }
     child[concept color=light-gray] { node[concept] (TeXweltQA)
        {Time Series} }
    }
    }
    child[concept color=light-gray] {
      node[concept] (TeXdoc)
        {Context}
        [clockwise from=284]
        child { node[concept] {Model-in-the-Loop }}
        child { node[concept] {Software-in-the-Loop}}
         child { node[concept] {Hardware-in-the-Loop}}
        }
    child[concept color=light-gray] {
      node[concept] (TeXdoc)
        {Scope}
        [clockwise from=219]
        child { node[concept] {Local Search}}
        child { node[concept] {Global Search}}
        }
    child[concept color=light-gray] {
      node[concept] (Blogs) {Eval. Metrics} [clockwise from=196]
      child { node[concept] {Number of Iterations}}
      child { node[concept] {Execution Time} }
      child { node[concept] (Cookbook)
        {Estimated Execution Time}}
    };
\end{tikzpicture}
\caption{A Taxonomy for Surrogate-Assisted Search-Based Testing (SA-SBT).}
\label{fig:taxonomy}
\end{figure}
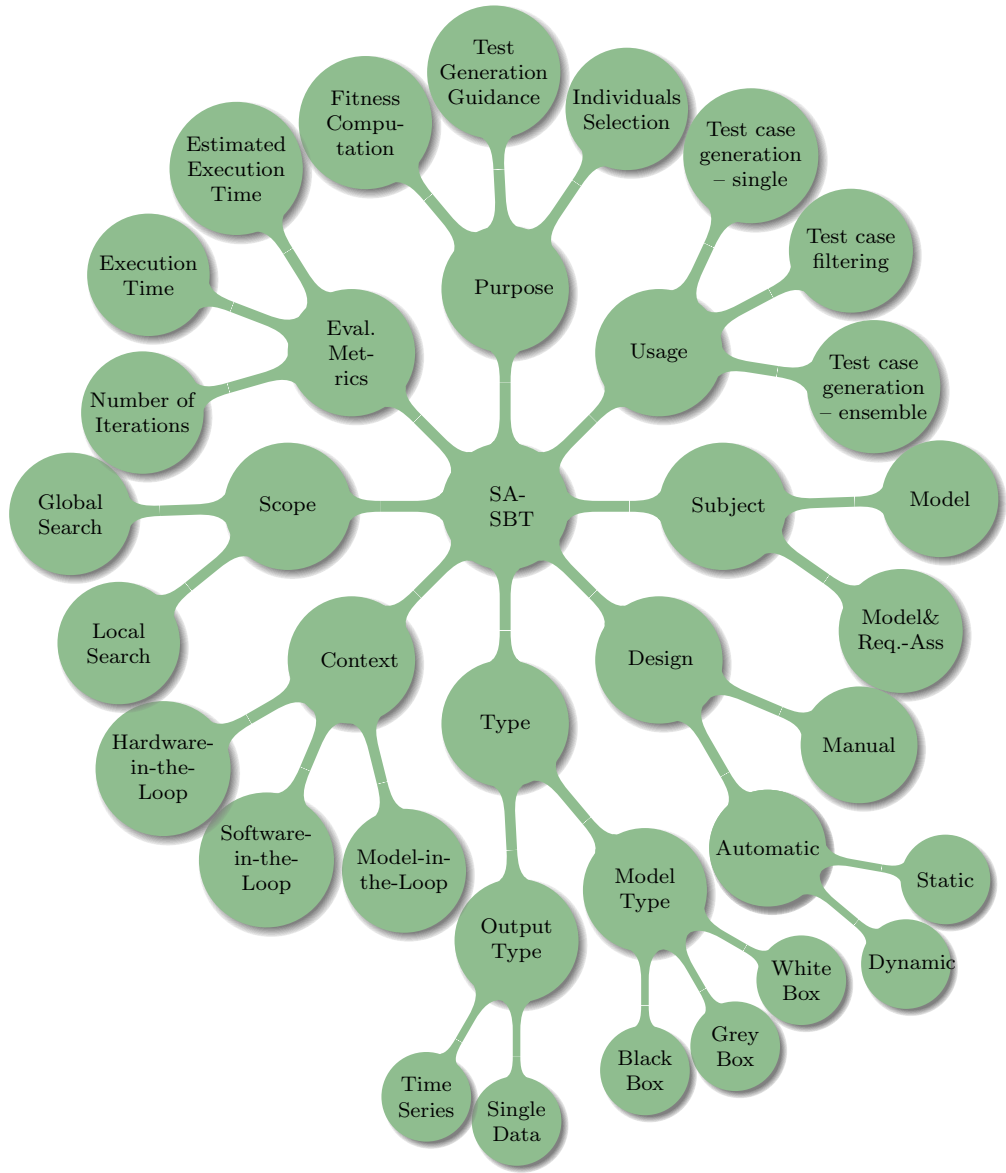

This section presents our taxonomy for SA-SBT research and classifies the existing research along these categories. Figure~\ref{fig:taxonomy} presents a mind map diagram representing our taxonomy. The diagram visualizes the concepts captured by our taxonomy and their relations. SA-SBT (the center node of the mind map diagram) is connected to eight categories representing relevant problems to be considered in the design of SA-SBT techniques (purpose, usage, subject, design, type, context, scope, and evaluation metrics). Each problem is connected with  sub-categories nodes representing relevant solutions and parameters to be considered when addressing these problems. In the following, we first describe a brief background on search-based testing (SBT). We then describe the categories of our taxonomy and discuss the sub-categories related to each of these categories.

\textbf{Background.} SBT relies on  metaheuristic search~\cite{Luke:13} to  generate a limited and effective set of test cases. Briefly, SBT algorithms work by building a set of initial candidate tests, iteratively applying tweak operations to one or more candidates to create new ones, and using a fitness function to decide whether to keep a candidate test for future iterations or to discard it.

\textbf{Purpose.} It refers to the way surrogates are combined with the metaheuristic search algorithm underlying an SBT technique and the search elements that they aim to improve. We identify three different purposes for surrogates: 
(1)~\emph{fitness computation}, (2)~\emph{test generation guidance}, and (3)~\emph{individuals selection}.

\begin{itemize}
    \item \emph{Fitness computation}. Surrogates reduce the fitness computation time by approximating the procedure for computing fitness values that often requires simulating the SUT.  
    \item \emph{Test generation guidance}. Surrogates  replace or augment search operators (a.k.a. tweak operators). Specifically, they help identify the regions of the search space that include fitter individuals. New candidates can thus be generated from within these regions. 
    \item \emph{Individuals selection}. Decisions during the search process  can be replaced with decisions guided by surrogates. For example, surrogates can be used to select fitter individuals for the initial population  at the start. 
\end{itemize}

\textbf{Usage.} It refers to the way surrogates improve the test generation approach of an SBT technique. We can use \emph{individual} surrogates or an \emph{ensemble} of surrogates. Surrogates may completely replace the SUT for test generation, or they may  filter poor candidate test cases, while the more promising candidates are evaluated on SUT.  We identify three different purposes for surrogates: 
(1)~\emph{test case generation using single surrogates}, (2)~\emph{test case filtering using single surrogates}, and (3)~\emph{test case generation using ensemble  surrogates}. 

\begin{itemize}
\item \emph{Test case generation using single surrogates}.  Surrogates are used to generate candidate test cases without any need to execute the original SUT during the test generation. To determine if the generated tests should be kept or discarded, they are assessed on the original SUT only at the end. 
\item \emph{Test case filtering using single surrogates}. Surrogates are used to prune candidates that have no chance of surviving the search selection process. They only discard poor test cases. The remaining test cases are assessed on the original SUT. 
\item \emph{Test case generation using ensemble  surrogates}. An ensemble of surrogates is used to assess individual test cases. A voting mechanism is used to determine, for a given test case, if the surrogates are precise enough, i.e., the surrogates produce the same (or close) fitness values. Otherwise, if the surrogates produce drastically different fitness values,  the test case is re-evaluated on the SUT.
\end{itemize}

\textbf{Subject.} It refers to the artifacts approximated by a surrogate model. We identify two kinds of artifacts to which surrogates can be applied:
\emph{model} and 
\emph{model$\&$requirement assessment approximation}.

\begin{itemize}
    \item \emph{Model}. The model approximation  technique relies on the creation of an approximation of the SUT that closely mimics its behavior, but is significantly cheaper to execute.
\item \emph{Model$\&$requirement-assessment}. It relies on a function to predict whether test cases satisfy or violate the system requirements.
That is, the surrogate approximates both the dynamics of the SUT and the assessment of the requirement.

\end{itemize}

\textbf{Design.} It refers to the process used to build a surrogate model.
We identify two processes for the design of surrogate models: \emph{manual} and \emph{automatic}. 

\begin{itemize}
    \item \emph{Manual}. In the manual process, engineers define a surrogate model by relying on their own experience and knowledge of SUT. 
\item  \emph{Automatic}. In the automatic process, engineers rely on some measurements to automatically construct a surrogate model. 
For example, these measurements can include logs recording the inputs and  outputs of the SUT. We consider two variations for the automatic process: \emph{dynamic} and \emph{static}.
The dynamic process improves surrogate models over time (e.g., thorough an approximation-refinement loop). 
The static process does not improve or modify surrogate models after they are computed. 

\end{itemize}

\textbf{Type.} It captures the characteristics of  surrogate models, and has two dimensions: the \emph{Model type} and the \emph{Output type}.

\begin{itemize}
    \item \emph{Model Type.} It captures the degree of information we have about the structure of surrogate models which can be one of the following three options: \emph{white box}, \emph{grey box}, and \emph{black box}. The white box type indicates that surrogates are designed manually based on the knowledge that engineers have about the SUT. Hence, there is full observability to their structure. Black box models are used when the structure of the surrogate model is unknown, and  engineers do not have any knowledge about the SUT, such as the physics and control laws that regulate the system behavior.
Gray box models are used when there is partial information about the structure of surrogate models. For example, gray box models are useful when the engineer knows the physics and control laws that regulate the system behavior but does not know some parameters, e.g., the heat transfer coefficient of a car engine.
Automatic techniques  are needed to compute black box models and unknown parameters of grey box models.
\item \emph{Output Type.} It refers to the format of the outputs of surrogate models.
We recognize two types of outputs: \emph{Single Data} and \emph{Time Series}.
Some surrogate models produce a single data point as output, e.g., a Boolean value indicating whether a property is satisfied or a real value indicating a degree of satisfaction. %Such data can be represented using variables of different types (e.g., Integer, Reals) or complex data structures. 
Alternatively,  surrogate models may generate time series data that is a sequence of data points indexed in time order. 
\end{itemize}

\textbf{Context.} 
It refers to the context where  surrogates are used. We identify three different usage contexts for surrogates:
\emph{Model-in-the-Loop (MiL)},
\emph{Software-in-the-Loop (SiL)}, and
\emph{Hardware-in-the-Loop (HiL)}.

\begin{itemize}
    \item \emph{Model-in-the-Loop}. It concerns with surrogate models that approximate a virtual and high-level representation of the system, e.g., some type of software model. 
    \item \emph{Software-in-the-Loop}. It relates to surrogate models that approximate software code.
    \item \emph{Hardware-in-the-Loop}. It is about surrogate models that are built based on the measurements obtained from the actual operating system including its  hardware components.
\end{itemize}

\textbf{Scope.} It refers to the portion of the search space surrogate models approximate.
The search domain a surrogate model targets can be \emph{global} or \emph{local}.

\begin{itemize}
    \item \emph{Global}. When surrogate models capture a global view of the search space of the SUT, they approximate the entire search space, and hence, enable an explorative search. 
    \item \emph{Local}. It refers to the case in which surrogate models capture a local view by approximating a region of the search space that includes desired information. In that case, surrogates are utilized for an exploitative search. 
\end{itemize}

\textbf{Evaluation Metrics.} It refers to the metrics used to assess an SBT approach. We identify three metrics: \emph{Number of Iterations}, \emph{Execution Time}, and \emph{Estimated Execution Time}. 

\begin{itemize}
    \item \emph{Number of Iterations}. It is the number of times an SBT approach executes the SUT. 
    \item \emph{Execution Time}. It measures the time the approach requires to detect a requirement violation.
\item \emph{Estimated Execution Time}. It predicts the time the approach would take to be applied to a compute-intensive system based on the results obtained for non-compute-intensive systems. 
\end{itemize}

We will refect on these metrics in detail in Section~\ref{sec:lessons}.

\section{Positioning against the State of the Art and Practice}
\label{sec:position}
To position the existing literature against the state of the arts and practice, one of the authors collected existing papers and classified relevant research literature along our taxonomy categories.
We considered 14 papers, the four subject papers of our study~\cite{Matinnejad:14,Abdessalem:16,Abdessalem:18,9283957} and a set of ten papers by other researchers that employ SBT for testing complex software systems identified by forward snowballing~\cite{1Beglerovic2017,3Arrieta2017,11Wang2022,12Innes2022,9Zhang2021,6Pedrielli2021,7Humeniuk2021,8Humeniuk2022,4Zhong2021,Ulhaq:22}, and classified relevant research literature along our taxonomy categories.
 We relied on Google Scholar for the forward snowballing: we identified new papers by consulting papers citing the four subject papers, we reviewed the retrieved papers, and we selected the papers that  proposed a new SBT approach. 
The forward snowballing activity lead to ten papers, including one paper~\cite{4Zhong2021} proposing fuzz testing techniques and six papers presenting falsification techniques~\cite{12Innes2022,9Zhang2021,6Pedrielli2021,7Humeniuk2021,8Humeniuk2022}. 
These works are specific instances of SBT techniques: 
fuzz testing is an SBT technique that injects unexpected inputs into a system to reveal software defects, 
and falsification techniques are SBT techniques driven by a requirement expressed using a formal language. 
We used these papers to demonstrate example research for different aspects of our taxonomy, and also, to analyze what sub-categories of our taxonomy are lesser studied compared to the others.  
Table~\ref{sec:classification} classifies relevant research literature along our taxonomy categories.
 Note that some papers are listed for some categories in Table~\ref{sec:classification}. 
 For example, for the evaluation metrics category, we do not list papers that do not evaluate their solutions. Table~\ref{sec:classification} shows that the SA-SBT techniques we reviewed cover all the nine top-level categories, and only two sub-categories (the Automatic-Static sub-category of the design category, and the SiL sub-category of the Context category) are not covered. In the following, we discuss the main findings for each category.

\begin{table}[th]
\footnotesize
\caption{Classifying the SA-SBT techniques based on our taxonomy}\label{sec:classification}
\vspace*{-.3cm}
\begin{center}
\scalebox{0.57}{\begin{tabular}{p{2.4cm}@{\hskip 0.05in} p{20cm}@{\hskip 0.5in} }
\toprule
\textbf{Category} &  \textbf{References} \\
\midrule
\multirow{3}{*}{\textbf{Purpose}} &  
\textbf{Fitness Computation}~\cite{Abdessalem:16,Matinnejad:14,Ulhaq:22,1Beglerovic2017,3Arrieta2017,6Pedrielli2021,8Humeniuk2022,11Wang2022}\\
 \cmidrule{2-2}
     & \textbf{Test GenerationGuidance}~\cite{Abdessalem:18,7Humeniuk2021,9Zhang2021}\\
 \cmidrule{2-2}
     & \textbf{Individual Selection}~\cite{9283957,4Zhong2021,12Innes2022} \\ 
\midrule
\multirow{3}{*}{\textbf{Usage}} &  
\textbf{TC generation--single}~\cite{9283957,Abdessalem:18, 1Beglerovic2017,3Arrieta2017,4Zhong2021,6Pedrielli2021,12Innes2022}\\
 \cmidrule{2-2}
     & \textbf{TC filtering}~\cite{Abdessalem:16,Matinnejad:14,9Zhang2021}\\
     \cmidrule{2-2}
     &  \textbf{TC generation--ensemble}~\cite{Ulhaq:22,7Humeniuk2021,8Humeniuk2022,11Wang2022}\\ 
\midrule
\multirow{2}{*}{\textbf{Subject}} & \textbf{Model}~\cite{9283957, 1Beglerovic2017,4Zhong2021,6Pedrielli2021,9Zhang2021,12Innes2022}\\
 \cmidrule{2-2}
& \textbf{Model and Requirement-Assessment}~\cite{Ulhaq:22,Abdessalem:16,Abdessalem:18,Matinnejad:14,3Arrieta2017,7Humeniuk2021,8Humeniuk2022,11Wang2022}
\\
\midrule
\multirow{3}{*}{\textbf{Design}}     &  \textbf{Manual} \cite{3Arrieta2017,7Humeniuk2021}\\
\cmidrule{2-2}
     & \textbf{Automatic--Static} \\
     \cmidrule{2-2}
      & \textbf{Automatic--Dynamic} \cite{9283957,Ulhaq:22,Abdessalem:16,Abdessalem:18,Matinnejad:14,1Beglerovic2017,4Zhong2021,6Pedrielli2021,8Humeniuk2022,9Zhang2021,11Wang2022,12Innes2022}\\
\midrule
\multirow{3}{*}{\textbf{Model Type}} &  \textbf{White Box}~\cite{9283957,1Beglerovic2017,7Humeniuk2021,8Humeniuk2022,9Zhang2021,12Innes2022} \\
\cmidrule{2-2}
     & \textbf{Gray Box}~\cite{9283957,8Humeniuk2022}\\
\cmidrule{2-2}
      & \textbf{Black Box}~\cite{9283957,Ulhaq:22,Abdessalem:16,Abdessalem:18,Matinnejad:14,3Arrieta2017,4Zhong2021,6Pedrielli2021,8Humeniuk2022,11Wang2022}\\
\midrule
\multirow{2}{*}{\textbf{Output Type}} &  \textbf{Single Data}~\cite{Ulhaq:22,Abdessalem:16,Abdessalem:18,Matinnejad:14,3Arrieta2017,4Zhong2021,7Humeniuk2021,8Humeniuk2022,11Wang2022}\\
\cmidrule{2-2}
     & \textbf{Time Series}~\cite{9283957,1Beglerovic2017,6Pedrielli2021,9Zhang2021,12Innes2022}\\
\midrule
\multirow{3}{*}{\textbf{Context}} &  \textbf{MiL}~\cite{9283957,Ulhaq:22,Abdessalem:16,Abdessalem:18,Matinnejad:14,1Beglerovic2017,3Arrieta2017,4Zhong2021,6Pedrielli2021,7Humeniuk2021,8Humeniuk2022,9Zhang2021,11Wang2022,12Innes2022}\\
\cmidrule{2-2}
     & \textbf{SiL}  \\
     \cmidrule{2-2}
    & \textbf{HiL} ~\cite{1Beglerovic2017}\\
\midrule
\multirow{2}{*}{\textbf{Scope}} &  \textbf{Global Search} ~\cite{9283957,Ulhaq:22,Abdessalem:16,Abdessalem:18,1Beglerovic2017,3Arrieta2017,4Zhong2021,6Pedrielli2021,7Humeniuk2021,8Humeniuk2022,9Zhang2021,11Wang2022,12Innes2022}\\
\cmidrule{2-2}
     & \textbf{Local Search}~\cite{Ulhaq:22,Matinnejad:14,6Pedrielli2021,11Wang2022}\\
\midrule
\multirow{3}{*}{\textbf{Metric}} &  
\textbf{Estimated Execution  Time}~\cite{9283957}\\
\cmidrule{2-2}
     & \textbf{Execution Time}~\cite{Abdessalem:16,Matinnejad:14,Ulhaq:22,Abdessalem:18,1Beglerovic2017,3Arrieta2017,7Humeniuk2021,8Humeniuk2022}\\
     \cmidrule{2-2}
     & \textbf{Num of Iterations}~\cite{9283957,4Zhong2021}\\     
\bottomrule
\arrayrulecolor{white}
\midrule
\end{tabular}}
\end{center}
\vspace*{-.6cm}
\end{table}

%Selected papers: 
%Search-based \cite{1Beglerovic2017,3Arrieta2017,6Pedrielli2021,7Humeniuk2021,8Humeniuk2022,11Wang2022,Ulhaq:22}
%Falsification \cite{12Innes2022,9Zhang2021,6Pedrielli2021,7Humeniuk2021,8Humeniuk2022,10Zhang2022}
%fuzzing: \cite{4Zhong2021,5She2019}

\textbf{Purpose}. Most of the techniques we reviewed use surrogates for fitness computation  (8 out of 14), followed by test generation guidance  (3 out of 14) and individual selection (3 out of 14).
Although research literature has focused less on using surrogates for test generation guidance and individual selection compared to fitness computation, incorporating surrogates can greatly enhance SBT techniques by pinpointing critical regions of the test input domain and selecting the most crucial individuals (test cases). 

%(FUZZING IS MORE ABOUT THE REST AND SBST ABOUT THE FIRST). 

\textbf{Usage}. Most of the techniques we reviewed use test case generation using single surrogates (7 out of 14), followed by test case generation using ensemble surrogates (4 out of 14), and test case filtering using single surrogates  (3 out of 14). 
Although techniques that use single surrogates for test case generation are prevalent, surrogate models are also used for other testing activities (i.e., test case filtering) and aggregated into ensembles. 
Since surrogate models are a relatively new technology in SBT, their potential usages are still unclear and will grow over time. 

\textbf{Subject}. For 6 out of 14 papers, the subject of the surrogate model is the original model under test, while for the remaining eight papers, it includes both the model and the requirement assessment. 
Using a surrogate model to approximate the original model under test and a fitness function to evaluate the different test cases ensures that the system's dynamics are approximated while the fitness computation is accurate.
Differently, also considering an approximation of the fitness function introduces additional noise.
There is a need for more empirical studies to precisely identify the situations where one of these approaches is preferred. 

\textbf{Design}. 
Most of the techniques we reviewed (12 out of 14) automatically compute the surrogate model and use dynamic processes that improve the surrogate model over time. 
Two techniques manually defined the surrogate model.
None of the techniques we reviewed used a static surrogate model, i.e., a fixed surrogate model that is not improved over time. 
Although automatic techniques do not require manually defining the surrogate model, in many applications, the manual design offers several benefits since engineers can incorporate knowledge about the structure of the SUT in the design of the surrogates. 
More empirical studies are needed to identify the scenarios in which one technique must be preferred over the other. 
%Our taxonomy can help understand and classify the differences and relationships between these techniques.

\textbf{Model Type}.
Most of the techniques we reviewed are black box (8 out of 14), followed by white box (4 out of 14). Two techniques,~\cite{9283957,8Humeniuk2022}, are generic and rely on other technologies such as system identification (SI)~\cite{bittanti2019model,ams1989system} that enable them to support black box, white box, and gray box model types.
In particular, SI enables us to create surrogates that belong to all the three categories. 
Although less studied, gray box technologies can offer a good trade-off in situations in which engineers have some knowledge about the system's internal structure and need to be aware of the values of some of the parameters that still need to be discovered.
%As done in other fields, e.g., in SI, it is necessary to identify the usage assumptions of different techniques and to provide guidelines for their selection.

\textbf{Output Type}. For 9 out of 14 techniques,  the output type is Single Data. 
The remaining five techniques use time series.
Therefore, in most cases, the surrogates directly predict the single data fitness value for a test input. 
Surrogates that use time series as output type are more recent~\cite{9283957}. Time series data leads to the generation of more data points that help with better training of surrogate models. 

\textbf{Context}. 
For all the techniques we considered, the surrogate context is MiL. One technique~\cite{1Beglerovic2017} claims that their approach  also supports HiL in addition to MiL.
The prevalence of MiL techniques is justified by the fact that MiL relies on simulators and system models, which are often publicly available. 
In contrast, SiL and HiL involve running the software under test on real or emulated hardware and require direct access to the source code and system hardware, which can be challenging in an industrial context. To overcome this hurdle, close collaboration with industry partners is often necessary. Furthermore, creating the necessary hardware infrastructure and setup can be both costly and time-consuming, particularly in an academic setting.
%In particular, in the industry context, these artifacts can only be obtained through close collaboration with industry, and non-disclosure agreements regulate their access. 
%Also, system code and hardware may be treated as proprietary. 
%We believe that MiL techniques are prevalent since the development and assessment of SiL and HiL require accessing artifacts that are usually not open-source.

\textbf{Scope}. The scope of 10 out of 14 techniques is global search. The scope of one techniques is exclusively local search. Three techniques consider both for local and global search.
We noticed that the boundary between global and local search was often blurry.
Additional research work is needed to precisely identify and define the boundary between the local and global search for SBT.

\textbf{Evaluation Metric}. Most of the techniques use execution time as an evaluation metric (8 out of 14), followed by the number of iterations (2 out of 14), and the estimated execution time (1 out of 14). 
Some techniques either did not use any evaluation metric, as they propose theoretical solutions which are not empirically evaluated, or their evaluation did not involve a comparison of SA-SBT solutions.  
Although the execution time measures the actual time required for fault-finding, it makes experiments hardly reproducible as it depends from the actual hardware platform on which the experiments are executed.
On the contrary, the number of iterations enables experiment replication as the hardware platform does not influence it. 
However, the execution time of the iterations may differ across different techniques.
We provide additional reflections and considerations on the evaluation metric category in Section~\ref{sec:lessons}.

\section{Replication Studies}
\label{sec:replications}
Papers listed in Table~\ref{sec:classification} do not present studies that try to replicate the approaches of the subject papers. 
In the next two sections, we present two replication studies based on two of the subject papers.  Specifically, we report replication studies of  NSGAII-DT~\cite{Abdessalem:18} and ARIsTEO~\cite{9283957}.  NSGAII-DT and ARIsTEO  belong to different sub-categories under the purpose, subject, model type, output type, and evaluation metric categories in our taxonomy. Replicating these complementary approaches allows us to  gain insightful observations and lessons learned.

\subsection{Replication Study: ADAS Case Study}
\label{sec:replicate}
%The papers we reviewed in Table~\ref{sec:classification} share some common aspects with the SBT approaches proposed in the subject papers.
%However, 
In this section, we report on the replication of the experiments from the subject paper proposing the  NSGAII-DT technique, which was evaluated using an industrial ADAS case study~\cite{Abdessalem:18}.
The replication attempt was performed by different researchers (the second and third authors of the paper) and by considering an ADAS which has been developed in collaboration with industrial partners and is different from that used in the original paper~\cite{Abdessalem:18}. 
In the following, we first describe the experiment setup of the original and the replication studies, then we present our results.

\subsubsection{Experiment setup}
\begin{figure}[t]
    \centering
\includegraphics[width=.7\textwidth]{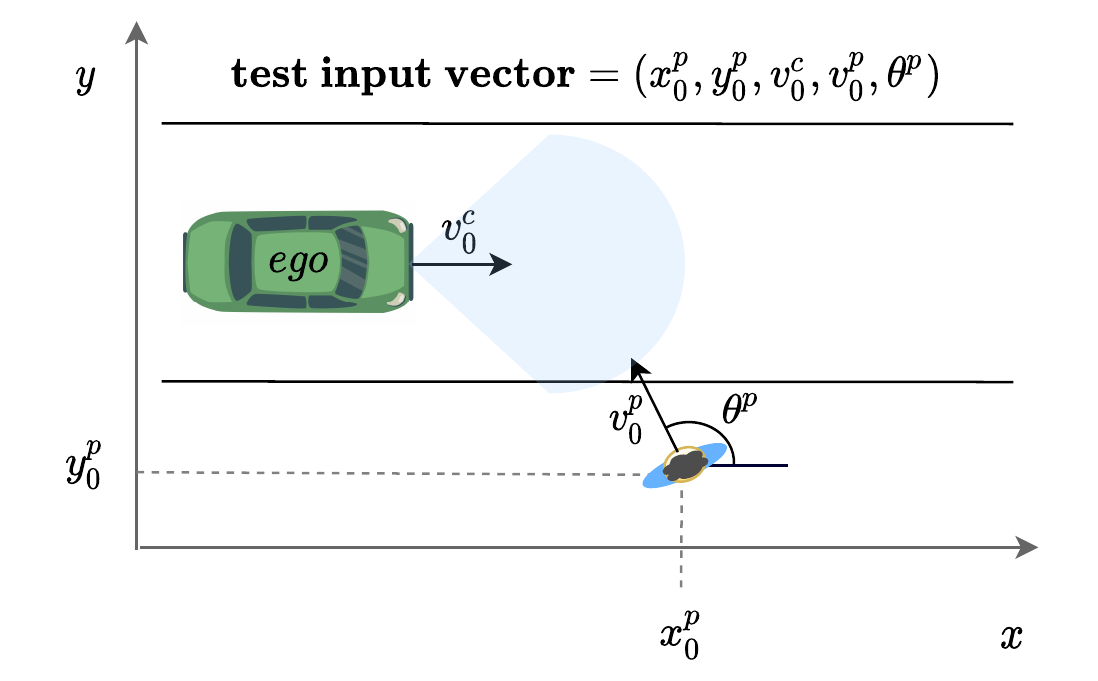}
    \caption{SUT of the original study~\cite{Abdessalem:18}: a pedestrian crosses the lane of the ego car equipped with the AEB system.}
    \label{fig:scenario_original}
\end{figure}

\textbf{System under test (SUT)}.  The system under test for the original subject paper is the Automated Emergency Breaking (AEB) system presented in~\Cref{fig:scenario_original}. The AEB identifies pedestrians in front of a vehicle and avoids collision by applying the brake when necessary. The ego car (i.e., the self-driving car equipped with AEB) drives from the left-end on an urban street, and a pedestrian starts  crossing the street from the middle of the street. 

\begin{figure}[t]
    \centering
\includegraphics[width=0.7\textwidth]{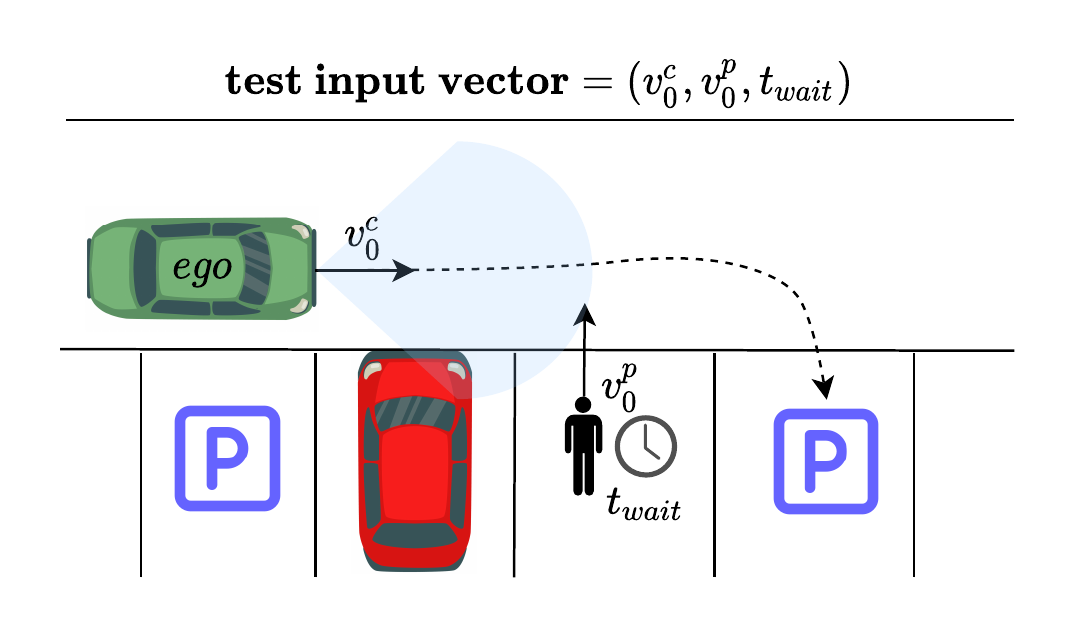}
    \caption{AVP testing scenario from the replication study: a pedestrian occluded by a parking car is crossing the lane of the ego car, which is approaching a free parking spot.}
    \label{fig:scenario_reflection}
\end{figure}

For the replication study,  we consider an Automated Valet Parking (AVP) system presented in~\Cref{fig:scenario_reflection}: A driver leaves the car at the entrance of a parking lot and AVP autonomously parks the car in an available parking spot. The ego car is equipped with a Lidar sensor and an AEB to ensure that the car does not collide with other objects, e.g., pedestrians or cars. The AVP is tested using a scenario where an occluded pedestrian appears in the path of the ego car as the ego car approaches a free parking spot. An occluded pedestrian is a pedestrian who is partially or completely obscured from the ego car's view by another object (e.g., a pillar or a parked vehicle).

%The SUT is equipped with an Automated Emergency Braking System (AEB), simple vehicle dynamics and a Trajectory Tracker to check that ego follows a precalculated path. The SUT components are provided as simulink models. We consider a scenario with two actors: an occluded pedestrian is crossing the lane of a moving ego (vehicle) at the parking lot.

\textbf{Search Space}. The original study tests are obtained by varying the following features of the AEB test scenario (see~\Cref{fig:scenario_original}): The speed of the ego car ($v_0^c$), the initial position of the pedestrian with respect to the ego car  ($x_0^p$, $y_0^p$), the speed ($v_0^p$) and the orientation ($\theta^p$) of the pedestrian, the weather type, which can be either normal, rainy and snowy, and the road shape, which can be either straight, curved or ramped. 
For the curved and ramped roads, the test input includes the curve radius and the ramp height.  
For the snowy and rainy weather types, the test input includes the level of precipitation and the presence of fog with different density levels. 

The search space for the replication study varies the speed of the ego car ($v_0^c$), the speed of the pedestrian ($v_0^p$) and  the time that  the pedestrian starts running after the start of the simulation ($t_{\mathit{wait}}$).  The variable $t_{\mathit{wait}}$ has a similar purpose as the variables ($x_0^p$, $y_0^p$) in the original study. Both $t_{\mathit{wait}}$ and ($x_0^p$, $y_0^p$) help generate different tests where the ego car and the pedestrian appear in different proximities to one another. The AVP testing excludes road shape and weather-related factors, as they are not essential for assessing AVP functionality. Given the confined parking lot environment, fluctuations in these conditions are considered non-essential for evaluating system performance.
In addition,  the AVP testing scenario is focused on occluded pedestrians, and for them, the initial position and the orientation  are not relevant since they are blocked from the ego car's field of view at the beginning. Instead, for occluded pedestrians, the relevant information for testing is when they appear in front of the ego car (variable $t_{\mathit{wait}}$), which is included as a test input in the replication study.  

%defined by three parameters: $ego\_velocity$ , the velocity which ego should achieve in the simulation, $pedestrian\_velocity$, the velocity which the pedestrian should accelerate to in the simulation, and the $pedestrian\_start\_walking$, the time when the pedestrian starts walking in front of the ego vehicle. In contrast to the original study in the reflection study we do not vary environmental parameters as weather or ramping height, since our SUT does not include camera or sensor devices whose performance is affected by these parameters.

%We use the following ranges for the search parameters: $ego\_velocity \in$ [0.1m/s, 5 m/s], $pedestrian\_velocity \in$ [0.1m/s, 5m/s], $pestrian\_start\_walking \in$ [0, 5s]. The ranges are defined following european regulations for driving a car in a parking lot and studies on pedestrian motion speed \cite{}.

Both the original and the replication studies use 
the PreScan simulator~\cite{prescan} to test their respective SUT.  Both studies set $100$ as the sampling rate for PreScan and generate simulations with the duration of $10$s. 

\textbf{Fitness functions.} Both studies define multiple fitness functions to assess the quality of individual simulations. 
Fitness functions measure how close the scenario is to a collision. The original study defines three fitness functions: ($\texttt{F}_1$)~The minimum distance between the pedestrian and the field of view of the ego car, ($\texttt{F}_2$)~The velocity of the ego vehicle at the time where the distance between the ego car and the pedestrian is at its lowest value, i.e., when the minimum for $\texttt{F}_1$ is computed, and ($\texttt{F}_3$)~The certainty of the detection of a pedestrian. By minimizing  $\texttt{F}_1$ and maximizing $\texttt{F}_2$, we obtain scenarios that either represent near-collision situations or collisions with a high speed of the ego car. In addition, the original study considers maximising $\texttt{F}_3$ since the study was focused on identifying scenarios where AEB detects a pedestrians but fails to avoid a collision with the pedestrian. Otherwise, scenarios where the pedestrian cannot be detected are due to the failure of the object detection component and not the failure of AEB.

The replication study defines two fitness functions: (1)~$\texttt{F}'_1$ (similar to $\texttt{F}_1$) this function  computes the  minimum distance between the ego car and the pedestrian. However, in the replication study, this distance is defined to be the distance between the pedestrian and the front side of the car. Hence, the $\texttt{F}'_1$ function is defined such that it assigns more optimal values to collisions where the pedestrian hits the car from the front. This is to provide more effective guidance for the search to generate scenarios where the collision is realistic and meaningful. That is,  the front side of the ego car hits a pedestrian. (2)~$\texttt{F}'_2$ (similar to $\texttt{F}_2$)  this function captures the velocity of the ego vehicle at the time where the distance between the ego car and the pedestrian is at its lowest value. The replication study does not consider the third fitness function (i.e., the certainty of detection) as in the original study, because the AVP does not provide detailed output that measures the certainty of detection. 
%WE NEED TO SAY WHY WE DID NOT HAVE F3 HERE (it is black box). 

%We consider that each simulation run produces a simulation output $S$, including position and velocity traces for each actor. The first fitness function $\mathcal{F}_1(S)$ computes an adapted distance between ego and the pedestrian at each simulation step and outputs the maximum of the value for one simulation run. The output is in the range from 0 to 1 and needs to be minimized. The function takes into account beside the euclidean distance the relative position of the pedestrian to ego, i.e. whether the pedestrian is in front of the ego vehicle or on the right of the vehicle. This function enables to distinguish between front and rear collisions and yields a value \textit{close} to 1 if a front collision happens. The second fitness function $\mathcal{F}_2(S)$ computes the velocity of the ego vehicle at the point in time when the first fitness function value is minimal. The value of this function needs to be maximized.

In addition to fitness functions, both studies define when a scenario represents a failure of the SUT. For both studies, a failure is when the ego car hits the pedestrian  with a high speed.  Failure revealing scenarios are identified by setting a threshold on $\texttt{F}_1$ and $\texttt{F}_2$ (resp. $\texttt{F}'_1$ and $\texttt{F}'_2$ in the replication study) to indicate that the car and the pedestrian collide and the car has a high speed at the time of the collision. Since the SUTs in the original and the replication studies are different, the threshold values for what defines a failure in these two studies are different too. Hence, the thresholds in these studies were defined independently and based on the requirements of their respective SUTs.

%, we use in addition a criticality function to guide the search to identify critical scenarios/situations. A critical scenario is when a front collision happens during simulation and the velocity of the ego is not 0 at the time of the collision, i.e., $\mathcal{F}_1(S) > 0.99\ \wedge\ \mathcal{F}_2(S) > 0)$.

%TODO check argumentation why 0.99 and not 0.98?

% In addition, we have a criticality function to guide the search and tell us what is a fault. In both studies, the criticality function is based on thresholds on two functions. 

\textbf{Initial population}. The replication study uses Latin Hypercube Sampling \cite{McKay79LHS}, a quasi-random sampling strategy, to generate the initial population. The original study used combinatorial testing using the Pledge tool \cite{henard:13} to generate an initial population of diversified individuals. Given that the purpose of both strategies is to generate a diverse initial set of individuals, we believe the differences in generating the initial populations do not significantly impact the results of our study. 

\textbf{Search Algorithm}. Both the original and the replication study rely on NSGAII-DT. 
NSGAII-DT  uses machine learning decision tree models and multi-objective evolutionary search~\cite{Luke:13}. Decision tree models are developed based on the scenarios generated at intermediary search iterations. The models learn the characteristics of the critical test scenarios and identify critical regions in an input space (i.e., the regions of a test input space that are likely to contain more fault-revealing test cases and test cases that reveal more severe faults). The subsequent search iterations then focus on the critical regions, and select and evolve critical scenarios in those regions. When the search inside the critical regions is terminated, a new decision tree model is re-generated using the entire population. NSGAII-DT continues by iteratively
building decision trees followed by search iterations focused on critical
regions identified by the trees until the search time budget runs out.
The original and replication studies consider the same values for the search parameters, such as mutation and crossover rates.  The main parameters and setup elements of the replication and the original studies are listed in Table~\ref{fig:my_label}.

% We used NSGAII but modified to be guided by surrogate. 

% General description from the original study. 

%\textbf{Initial Sampling} For the sampling of the initial population we use Latin Hypercube Sampling \cite{X}, an improved random sampling strategy. In the original paper is used the Pledge tool \cite{} with the algorithm X. We deviate from this sampling strategy because...

%\textbf{Operators} 
% Mutation and cross-over (what are these in the original and replication studies.)
%For the replication we use Simulated Binary Crossover with probability 0.7 and eta 20 (in original study X). For the mutation of individuals as in the original paper we use the probability 1/n, where n is the number of search parameters, for the eta we use 15. The number of offsprings produced every generation equals to the population size (in the original study it is...). In contrast to the original study we do not have constraints on search parameter values, because we do not modify environmental parameters and positions of actors.

\begin{table}[t]
    \centering
    \footnotesize
\caption{Experimental setup for replication study  vs. the study in original paper~
\cite{Abdessalem:18}}\label{fig:my_label}
 \begin{tabular}[t]{l p{4.5cm} p{4.5cm}}
 \toprule
      \textbf{Category} & \textbf{Replication study setup} & \textbf{Original study setup} \\
      \midrule
      SUT  & Longitudinal/Latitudinal vehicle dynamics, Path Follower, Lidar  &  3D vehicle dynamics, Object detection, Lidar \\
      Sampling rate  & 1/100  & 1/100 \\
      Scenario & Pedestrian crossing street & Pedestrian crossing street \\
      \# fitness functions  & 2 & 3 \\
      Initial sampling & LHS & Combinatorial Testing \\
      Mutation probability & 1/n & 1/n \\
      Crossover probability & 0.6 & 0.6 \\
      \bottomrule
  \end{tabular}
\end{table}
%\textbf{Search configuration}

%How many runs do we run each of the algorithms. 
%What is the population size.
%What is the search time for one run.

\subsubsection{Results}
The original study provides two sets of metrics to evaluate the results of NSGAII-DT: (1)~Quality indicators for Pareto-based search algorithms~\cite{TaoParetoQI20}, and (2)~the number of distinct critical scenarios representing failures of the SUT.  As a Pareto-based search algorithm, NSGAII-DT aims to find a set of non-dominated solutions in a multi-objective optimization problem. These algorithms are typically assessed based on a special category of metrics, referred to as \emph{quality indicators}~\cite{TaoParetoQI20} that compare Pareto-based search algorithm in terms of convergence, uniformity, spread and cardinality.
In particular, the original study used three well-known quality indicators, Hypervolume (HV), Spread ($\Delta$)  and Generational Distance (GD), to compare the results of NSGAII-DT with those of NSGAII. Below we describe the experiment setup for our replication study and then compare and discuss the results of both studies.

\textbf{Setup.} We evaluate the performance of NSGAII-DT and NSGAII on the AVP case study for a minimum of $300$ minutes. We note that NSGAII served as a baseline in the original study.  The average execution time of a single AVP simulation is approximately 19 seconds, allowing for at least 1000 simulations (fitness evaluations) to be completed within the 300-minute computation time.  We rerun each of NSGAII-DT and NSGAII for 10 times to account for their randomness. The experiments were performed on a computer with an i7-8700 processor with 3.19 Ghz (12 cores) and 32 GB RAM.

Note that the AEB simulations in the original study took significantly longer than the AVP simulations, with an average execution time of 1.2 minutes for AEB and 19 seconds for AVP. In the original study, both NSGAII-DT and NSGAII algorithms were run for a 24-hour period, providing enough time for at least 1000 minimum AEB simulations, similar to our replication study.

%for minimal $1000$ evaluations with a population size of 100. The remaining parameters and the CERT algorithm for the DT computation is configured as in the original study. We cannot control the maximal number of evaluations for runs with NSGAII-DT since the number of evaluations done by NSGAII depends on the number and size of the the critical regions identified by the learned decision trees. We rerun each of the algorithm for 10 times instead of 15 times, due to the time budget.  The simulation takes longer in the original study since the AEB needs to process vision based data and because we do not vary environmental information.

\textbf{Metrics.} 
For NSGAII, we compute HV, Spread ($\Delta$)  and GD after each generation. But, for NSGAII-DT, we compute these metrics after each generation of each of the NSGAII algorithm runs performed inside the critical regions. HV measures the size of the space covered by the members of a Pareto front generated by a search algorithm~\cite{TaoParetoQI20}. The higher this size, the better the results of the algorithm. GD measures the Euclidean distance between members of the calculated Pareto front and the nearest solutions on a reference Pareto front~\cite{TaoParetoQI20}. The lower the value of GD, the more optimal the calculated Pareto front solutions are. Spread measures the extent of spread among the members of a Pareto front generated by a search algorithm~\cite{DebNSGA2_02}. The lower the Spread values, the better spread out the search outputs. In addition, we compute the number of distinct critical scenarios generated by NSGAII and NSGAII-DT, following the definition provided in the original study, where distinct scenarios are defined as those that differ in at least one input value~\cite{Abdessalem:18}.

\textbf{Data Availability.}  The implementation of NSGAII and NSGAII-DT for AVP and the detailed results of these algorithms for our replication study are available online~\cite{resultsReplication1}.

\begin{figure}[t]
    \centering
    \includegraphics[width=.85\textwidth]{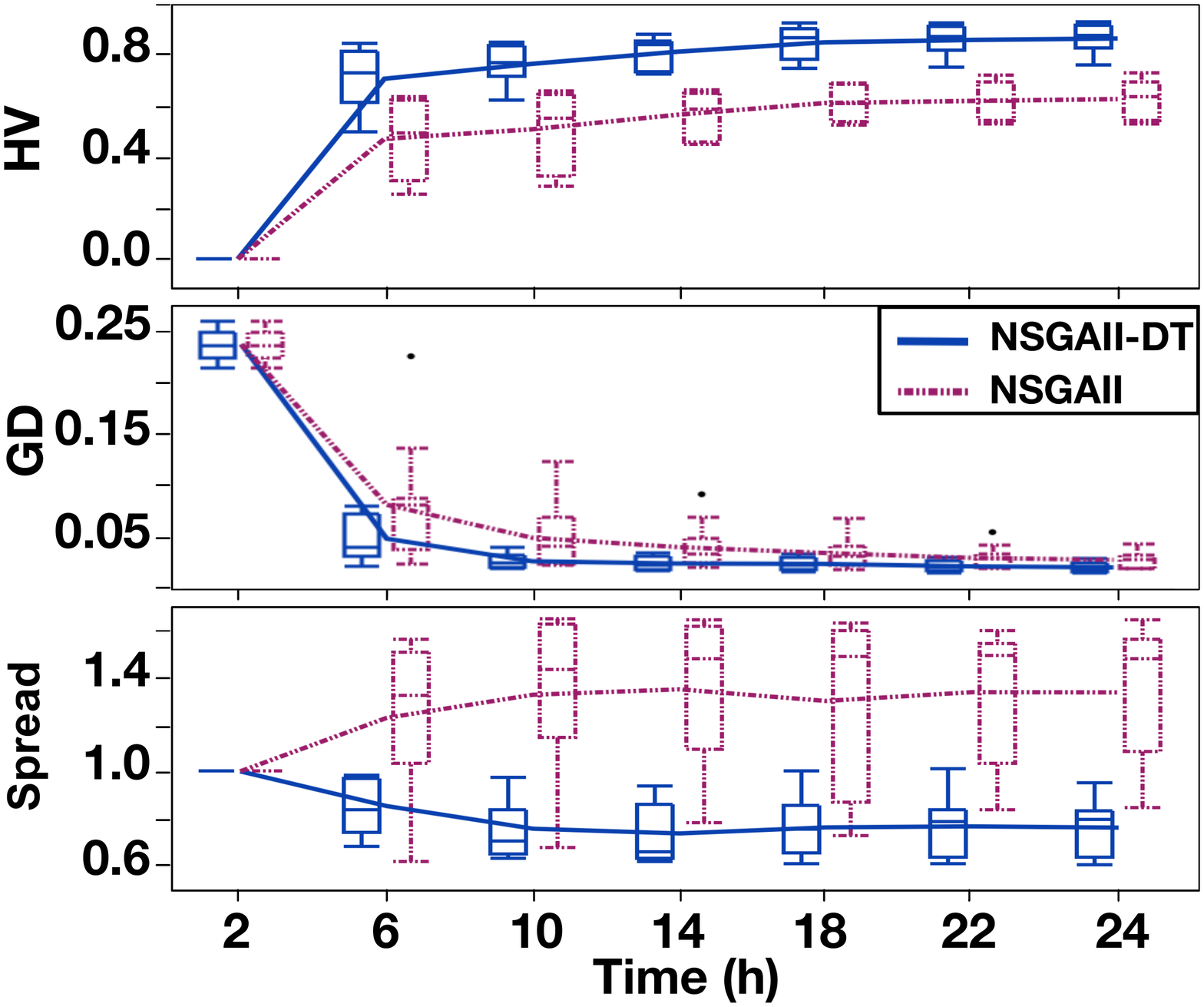}
    \caption{The comparison of  HV, GD and Spread ($\Delta$) values obtained by NSGAII and NSGAII-DT from the original study~\cite{Abdessalem:18}.}
    \label{fig:indicatorsOri}
\end{figure}

\begin{figure}[t]
    \centering
    \includegraphics[width=1\textwidth]{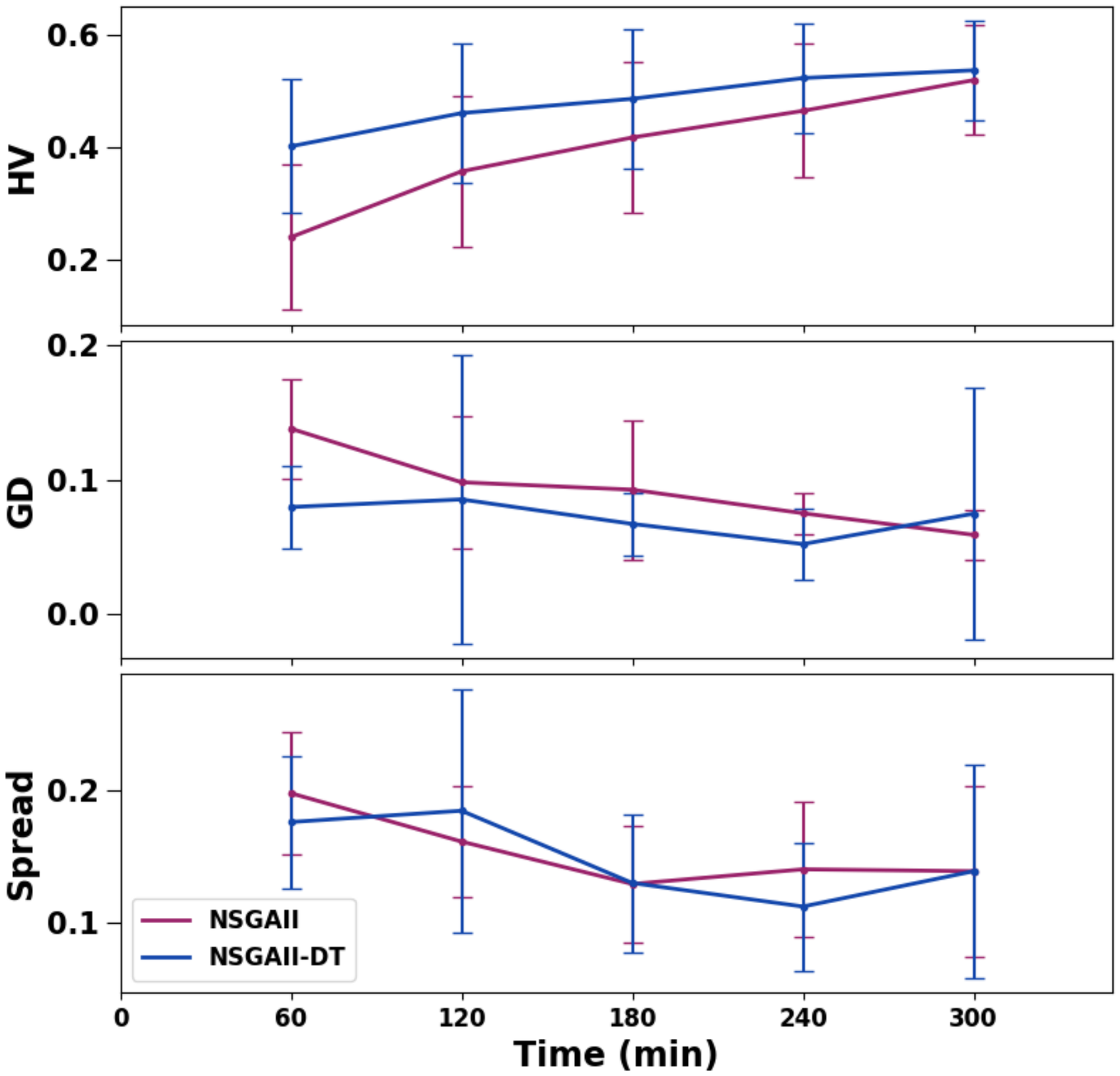}
    \caption{The comparison of  HV, GD and Spread values obtained by NSGAII and NSGAII-DT from the replication study.}
    \label{fig:indicatorsRep}
\end{figure}

\textbf{Comparison of the results.} 
Figures~\ref{fig:indicatorsOri} and \ref{fig:indicatorsRep} show the results of the HV, Spread and GD  metrics over time for NSGAII and NSGAII-DT for our original and replication study, respectively. The results for the original study are shown at  every four-hour time interval starting at 2h until the time limit of $24$h, while the results for the replication study are shown at every $60$ min starting from $60$min and until $300$min. Since the original and replication experiments are performed on different case studies with different inputs and fitness functions,  we cannot make direct comparisons between the values of these metrics, but instead, we aim to compare the trends of these metrics over time.

%Note that the metric results in Figure~\ref{fig:indicatorsRep} are obtained by plotting the metrics computed based on the results obtained up to a time instant irrespective of intermediary results from the critical regions which are still in progress.  every four-hour time interval starting at 2h as well as the results at the end of the search time limit. In contrast, the metric results in Figure~\ref{fig:indicatorsRep} by first plotting the detailed results obtained from each critical region including the intermediary results, and then using spline interpolation \cite{Interpolation} to calculate the metric values at different time intervals. This difference in plotting is due to the fact that NSGAII-DT does not maintain a fix population size and the notion of generation for NSGAII-DT is different from that of NSGAII. 

In both the original and replication study, the HV values for NSGAII-DT are better than those of NSGAII at the start of the experiments. However, over time, the differences between the HV values of the two algorithms tend to decrease. In the replication study, this reduction is quite steep, with the HV values of NSGAII and NSGAII-DT eventually becoming equal. In contrast, in the original study, the HV values of the two algorithms seem to stabilize at a certain point.  The GD values of NSGAII-DT are slightly better than those of NSGAII over the entire duration both in the original study and in the replication study. However, at the end of the simulation, which is 300 minutes, the GD values of NSGAII slightly improve over those of NSGAII-DT in the replication study. Further, the variations of the GD values for  NSGAII-DT in the replication study are noticeably higher than those of the GD values of NSGAII-DT in the original study.
There are remarkable differences between the replication and the original studies  in the values of the Spread metric obtained for NSGAII-DT. While in the original study the values of Spread for NSGAII-DT are considerably better than those of NSGAII, for the replication study the values of Spread for both NSGAII and NSGAII-DT are almost the same.

%up to the Xth evaluation. At evaluation X, Y and Z we the standard deviation is high compared to the standard deviation for the other number of evaluations, which is due to the fourth run \ref{figure} which notable deviates from the other runs ((explain how)). In contrast to the original study, at the last iteration the GD value (Z) of NSGAII-DT is higher then the one of NSGAII (W). The difference between both GD values is similiar to the one in the original study.

Table~\ref{tbl:results_critical} shows the number of distinct critical scenarios generated by NSGAII-DT and NSGAII in the replication and original studies. As shown in the table, in both studies, provided with the same search time budget, NSGAII-DT is able to find considerably more distinct critical scenarios compared to NSGAII. In particular, the number of distinct critical scenarios that NSGAII-DT finds are, respectively,  $1.8$ and $2.7$ times more than those that NSGAII can find in the original and replication studies.

\begin{table}[t]
\centering
    \footnotesize
\caption{The number of distinct critical scenarios obtained by NSGAII and NSGAII-DT from the replication and original study.}
    \label{tbl:results_critical}
    \centering
\begin{tabular}{c c c}
\toprule
    & \textbf{Replication study} & \textbf{Original study} \\
\midrule
      NSGAII & 639 & 411 \\
     NSGAII-DT  & 1733 & 731 \\
Ratio NSGAII-DT / NSGAII &  2.7  & 1.8 \\
\bottomrule
    \end{tabular}
\end{table}

\textbf{Summary.} The replication study supports some of the conclusions of the original study. However, there are some disparities with some specific findings. Both studies confirm that the surrogate-based SBT method, NSGAII-DT, generates significantly more distinct failures than its non-surrogate counterpart  within the same search time budget.  However, the consistency of the results comparing the quality of the Pareto-based solutions generated by NSGAII and NSGAII-DT is questionable. The HV and GD results from the replication study partially align with those from the original study, whereas the results  of the Spread metric are significantly different. Despite our efforts to align the two studies as much as possible,  the SUTs, and hence, the test inputs and fitness functions remain to be different. These factors may have contributed to the discrepancies in the outcomes. Independently from the real sources of these differences, our replication study allowed us to gain a deeper understanding of the limitations and significance of the metrics used to evaluate NSGAII-DT, which we elaborate on in Section~\ref{sec:lessons}.

\subsection{Replication Study: Simulink}
\label{sec:simulink}
In this section, we report on an attempt to replicate the experiments of \ARIsTEO, an approximation-refinement testing technique driven by surrogate models proposed by one of the subject papers~\cite{9283957}.
The \ARIsTEO technique is showcased by a plugin of \STaLiRo~\cite{S-Taliro}, a falsification-based testing tool for Simulink models.
In their paper, the authors of \ARIsTEO, evaluated their contribution by considering a set of case studies which included some of the benchmark models of the ARCH competition~\cite{DBLP:conf/arch/ErnstABCDFFG0KM21} and one additional industrial case study from the satellite domain. 
The industrial case study is not publicly available since the authors could not release it due to a non-disclosure agreement.
However, the authors provided a complete replication package with the other benchmark models and results. 
In addition, they participated in the ARCH competition, editions of 2020~\cite{DBLP:conf/arch/ErnstABDFFMMPPY20}, 2021~\cite{DBLP:conf/arch/ErnstABCDFFG0KM21}, and 2022~\cite{ARCH22:ARCH_COMP_2022_Category_Report}.
The ARCH competition is a friendly yearly competition between testing tools for continuous and hybrid systems~\cite{ARCHWEBSITE}.
The competition includes several participants, such as \FalCAuN{}~\cite{Waga20}, 
\STaLiRo~\cite{S-Taliro},
\falsify~\cite{yamagata2020falsification},
\FalStar~\cite{ErnstSZH2018-FalStar},
\foresee~\cite{falsQBRobCAV2021}.
The models and requirements considered by the competition are publicly available and provided under different licenses.

\begin{table}[ht]
	\centering
    \setlength{\tabcolsep}{6pt}
	\caption{Results for piecewise continuous input signals.
	}
	\label{table:piecewise}
 \footnotesize
 \resizebox{0.75\columnwidth}{!}{
    \begin{tabular}{l rrr rrr rrr }
			\toprule
%                & \multicolumn{9}{c}{\bf Year}\\
%			\midrule
                & \multicolumn{3}{c}{\bf2020}
			& \multicolumn{3}{c}{\bf2021}
			& \multicolumn{3}{c}{\bf2022}\\
               \midrule
			& \FR & \Savg & \Smed
			& \FR & \Savg & \Smed
			& \FR & \Savg & \Smed\\
			\midrule
			AT1  & 0 & - & - &  0 & - & - & 0 & - & - \\
			AT2  & 50 & 4.4 & 4 & 50 & 4.1 & 3 & 10 & 5.4 & 2.0\\
			AT51 & \multicolumn{3}{c}{nSbyA} & 0 & - & - & 0 & - & -\\
			AT52 & \multicolumn{3}{c}{nSbyA} & 50 & 3.2 & 2 & 10 & 1.7 & 1.5\\
			AT53 & \multicolumn{3}{c}{nSbyA} & 50 & 2.6 & 2 & 10 & 12.8 & 8.5\\
			AT54 & \multicolumn{3}{c}{nSbyA} & 3 & 295.1& 300 & 0 & - & -\\
			AT6a & 45 & 90.7 & 69 & 50 & 39.0 & 23 & 10 & 93.4 & 70.0\\
			AT6b & 50 & 18.1 & 15 & 49 & 101.7 & 83 & 4  & 112.3 & 127\\
			AT6c & 44 & 95.6 & 66 & 50 & 27.2 & 16 & 10 & 124.2 & 111.5\\
			AT6abc & \multicolumn{3}{c}{NPofB} & 50 & 27.2 & 16 & 10 & 73.4 & 30.5 \\
			\cmidrule(r){1-1} \cmidrule(lr){2-4} \cmidrule(lr){5-7} \cmidrule(lr){8-10} 
			NN  & 50 & 62.8 & 46 & 50 & 62.8 & 46 & 1 & 299.0 & 299.0\\
   NNx  &  \multicolumn{3}{c}{NPofB} & \multicolumn{3}{c}{NPofB} & 0 & - & - \\
			\cmidrule(r){1-1} \cmidrule(lr){2-4} \cmidrule(lr){5-7} \cmidrule(lr){8-10} 
			WT1 & 50 & 15.6 & 10 & 50 & 13.0 & 10 & \multicolumn{3}{c}{NPofB}\\
			WT2 & 50 & 1.5 & 1 & 50 & 1.3 &1  & \multicolumn{3}{c}{NPofB} \\
			WT3 &50 &  3.4 & 3 & 50 & 2.4 & 2  & \multicolumn{3}{c}{NPofB}\\
			WT4 &50 & 1.0 & 1 & 50 & 1.0 & 1  & \multicolumn{3}{c}{NPofB}\\
			\cmidrule(r){1-1} \cmidrule(lr){2-4} \cmidrule(lr){5-7} \cmidrule(lr){8-10} 
			CC1 &50 & 16.1& 11 & 50 & 27.0 & 15 & 10 & 24.8 & 17.5\\
			CC2 &50 & 1.0 & 1 & 50 & 9.2 & 6  & 10 & 14.6 & 9.0\\
			CC3 &50 & 45.8 & 27 & 50 & 56.3 & 46 & 10 & 29.5 & 35.0\\
			CC4 & 50 & 1.0 & 1 & 0  & - & -& 0 & - & -\\
			CC5 & 49 & 52.5 & 40 & 50 & 25.9 & 17 & 10 & 18.8 & 18.0\\
			CCx &  \multicolumn{3}{c}{NPofB} & 15 & 250.0 & 300 & 4 & 134.0 & 74.0\\
			\cmidrule(r){1-1} \cmidrule(lr){2-4} \cmidrule(lr){5-7} \cmidrule(lr){8-10} 
			F16 & \multicolumn{3}{c}{nSbyA} & 0 & - &- & 0 & - &-  \\
			\cmidrule(r){1-1} \cmidrule(lr){2-4} \cmidrule(lr){5-7} \cmidrule(lr){8-10} 
			SC  &50 & 1.0 & 1 & 0 & - & - & 0 & - & -\\
			\bottomrule
	\end{tabular}}
\end{table}

\begin{table}[t]
	\centering
 \footnotesize
    \setlength{\tabcolsep}{6pt}
	\caption{Results for constrained input signals.
	}
	\label{table:constrained}
 \footnotesize
 \resizebox{0.75\columnwidth}{!}{
    \begin{tabular}{l rrr rrr rrr }
			\toprule
%                & \multicolumn{9}{c}{\bf Year}\\
%			\midrule
                & \multicolumn{3}{c}{\bf2020}
			& \multicolumn{3}{c}{\bf2021}
			& \multicolumn{3}{c}{\bf2022}\\
               \midrule
			& \FR & \Savg & \Smed
			& \FR & \Savg & \Smed
			& \FR & \Savg & \Smed\\
			\midrule
			AT1  & 0 & - & - &
                        0 & - & - & 
                        0 & - & - \\
			AT2  & 50 & 4.5 & 3 
                        & 50 & 5.1 & 4 
                        & 10 & 12.9 & 8.0\\
			AT51 & \multicolumn{3}{c}{nSbyA} 
                & 50 & 8.4 & 6 
                & 10 & 19.0 & 10.0\\
			AT52 & \multicolumn{3}{c}{nSbyA} 
                & 50 & 3.5 & 3 
                & 9 & 74.7 & 46.0\\
			AT53 & \multicolumn{3}{c}{nSbyA} 
                & 50 & 2.6 & 2 
                & 10 & 1.4 & 1.0\\
			AT54 & \multicolumn{3}{c}{nSbyA} 
                & 50 & 29.4 & 17 
                & 10 & 44.0 & 40.0\\
			AT6a & 41 & 116.3 & 72 
                    & 47 & 103.1 & 80 
                    & 7 & 65.1 & 62.0\\
			AT6b & 50 & 36.3 & 27 
                & 45 & 164.7 & 174 
                & 5  & 91.2 & 74.0\\
			AT6c & 44 & 89.8.6 & 73 
                & 49 & 89.1 & 60 
                & 7 & 175 & 141.0\\
			AT6abc & \multicolumn{3}{c}{NPofB} 
            & 50 & 79.1 & 70 
            & 9 & 84.1 & 81.0 \\
			\cmidrule(r){1-1} \cmidrule(lr){2-4} \cmidrule(lr){5-7} \cmidrule(lr){8-10} 
   AFC27  & 50 & 2.3 & 1 
   & 50 & 2.3 & 1 
   & 9 & 21.9 & 25.0\\
    AFC29  & 50 & 28.5 & 23 
    & 50 & 1.0 & 1 
    & 10 & 3.4 & 3.5\\
    AFC33  & 50 & 24.7 & 16 
    & 50 & 1.0 & 1 
    & 0 & - & -\\
   \cmidrule(r){1-1} \cmidrule(lr){2-4} \cmidrule(lr){5-7} \cmidrule(lr){8-10}
			NN  & 50 & 62.8 & 46 
   & 50 & 62.8 & 46 
   & 10 & 117 & 82.5\\
   NNx  &  \multicolumn{3}{c}{NPofB} & 50 & 1.0 &1 & 0 & - & - \\
			\cmidrule(r){1-1} \cmidrule(lr){2-4} \cmidrule(lr){5-7} \cmidrule(lr){8-10} 
			WT1 & 50 & 1.4 & 1 
                & 50 & 1.4 & 1 
                & \multicolumn{3}{c}{NPofB}\\
			WT2 & 50 & 1.0 & 1 
                & 50 & 1.0 & 1  
                & \multicolumn{3}{c}{NPofB} \\
			WT3 & 50 &  1.1 & 1 
           & 50 &  1.1 & 1  
            & \multicolumn{3}{c}{NPofB}\\
			WT4 & 50 & 1.0 & 1 
            & 50 & 1.0 & 1  
            & \multicolumn{3}{c}{NPofB}\\
			\cmidrule(r){1-1} \cmidrule(lr){2-4} \cmidrule(lr){5-7} \cmidrule(lr){8-10} 
			CC1 & 50 & 28.8 & 22 
                & 50 & 9.1 & 6 
                & 10 & 9.1 & 8.0\\
			CC2 & 50 & 1.0 & 1 
                & 50 & 6.7 & 4  
                & 10 & 10.8 & 9.0\\
			CC3 & 50 & 18.1 & 16 
        & 50 & 18.9 & 15 
        & 10 & 12.8 & 13.4\\
			CC4 & 50 & 1.0 & 1 
            & 0  & - & -
            & 0 & - & -\\
			CC5 & 48 & 66.9 & 44 
            & 50 & 29.1 & 12 
            & 10 & 21.1 & 11.5\\
			CCx &  \multicolumn{3}{c}{NPofB} & 20 & 229.4 & 12 
   & 3 & 97 & 103.0\\
			\cmidrule(r){1-1} \cmidrule(lr){2-4} \cmidrule(lr){5-7} \cmidrule(lr){8-10} 
			\cmidrule(r){1-1} \cmidrule(lr){2-4} \cmidrule(lr){5-7} \cmidrule(lr){8-10} 
			SC  &50 & 1 & 1 & 0 & - & - & 0 & - & -\\
			\bottomrule
	\end{tabular}}
\end{table}

We report on our attempt to replicate the results generated from ARIsTEO from editions
2020~\cite{DBLP:conf/arch/ErnstABDFFMMPPY20},
2021~\cite{DBLP:conf/arch/ErnstABCDFFG0KM21},
and 2022~\cite{ARCH22:ARCH_COMP_2022_Category_Report}.

\subsubsection{Experiment Setup}
Our benchmark consists of seven models: Automatic Transmission (AT)~\cite{ARCH14}, Fuel Control on Automotive Powertrain (AFC)~\cite{JinHSCC2014}, Neural Network Controller (NN)\cite{NN}, Wind Turbine (WT)~\cite{ARCH16:Hybrid_Modelling_of_Wind}, Chasing Cars (CC)~\cite{hu2000towards}, Aircraft Ground Collision Avoidance System (F16)~\cite{ARCH18:Verification_Challenges_in_F_16}, and Steam Condenser (SC)~\cite{YaghoubiHSCC}.
These Simulink models have different sizes and complexity: the number of Simulink blocks spans from 13 (CC) to 302 (AFC). 
The models are from different domains: automotive  (AT, AFC), neural networks (NN), and energy (SC).
AFC is from Toyota. 
The models come with different licenses, including GPLv2.

The models of the ARCH competition are associated with $27$ requirements expressed using a logic-based language. 
Each requirement is associated with an identifier that starts with the model's name.
The rows of \Cref{table:piecewise} and \Cref{table:constrained} represent different requirements. 
For example, the row with the identifier AT6a refers to the requirement AT6a of the AT model. 
The interested reader can find additional information about the models and requirements in the report of the ARCH competition (i.e., \cite{DBLP:conf/arch/ErnstABCDFFG0KM21,DBLP:conf/arch/ErnstABDFFMMPPY20,ARCH22:ARCH_COMP_2022_Category_Report}).

The same configuration is considered for \ARIsTEO across the three years it participated in the competition.
Specifically, \ARIsTEO is configured as follows.
We used an \textsc{arx} model (\textsc{arx}-$2$) with order $na=2$, $nb=2$, and $nk=2$\footnote{\url{https://nl.mathworks.com/help/ident/ref/arx.html}}
as the structure for the surrogate model used in the approximation-refinement loop of \ARIsTEO.
For models with multiple inputs and outputs, the dimension of the matrix $na$, $nb$, and $nk$ are changed depending on the number of inputs and outputs.
For the other parameters, we assigned the default configuration values of \STaLiRo.
We considered the same parametrization of \STaLiRo\ for the input signals.
\ARIsTEO executes the original Simulink model to learn the initial surrogate model.  
The competition sets the cut-off values for the number of simulations of the original model to $300$ for $2020$ and $2021$ and  do not impose  a maximal number of simulations that can be run in one falsification trial for $2022$.
This choice was performed to enable a more accurate assessment of the difficulty of the benchmarks.
We considered this value also for the number of simulations of the surrogate model (per trial).
However, we considered $300$ as the maximum number of iterations for \ARIsTEO in $2022$ since we aimed to maintain consistency with the previous years of the competition. 
Notice that this choice penalizes \ARIsTEO over the other competing tools in $2022$.

During the competition, the participants must test the model for each requirement: \emph{piece-wise continuous input signals}, and \emph{constrained input signals}. 
The first option let the participant decide the shape of the input signals, and the second option precisely fixes the format of the
input signal. 
\Cref{table:piecewise} and \Cref{table:constrained} respectively report the results obtained by \ARIsTEO for the two options.
In addition, the competition required participants to execute experiments $50$ times in $2020$ and $2021$ and $10$ times in $2022$ due to the non-determinism of the underlying search-based algorithms.
In $2022$ the number of experiments was reduced from $50$ to $10$ since there was no upper bound on the maximal number of simulations that can be run in one falsification trial.

%We use Simulated Annealing, its default search algorithm of \staliro.
%We set the value \niterations for the maximum number of iterations  

\subsubsection{Results}
\Cref{table:piecewise} and \Cref{table:constrained} report the results of our experiments.
For each year, they report the falsification rate with respect to the number of trials (\FR),	mean (\Savg) and median (\Smed) number of simulations (rounded down) over successful trials (``--'' if \FR\ is zero).
Cells labeled with NPofB (Not Part of the Benchmark) specifies that the model was not part of the benchmark models considered by the competition that year.
For the first year (2020) of the competition, the cells for the benchmarks  AT51, AT52, AT53, and AT54 are labeled with nSbyA (not supported by \ARIsTEO) since \ARIsTEO could not process the benchmark.
Indeed, for the first year of the competition, \ARIsTEO could not support property specifications that use locations.
Locations are used by \STaLiRo~\cite{S-Taliro} (the tool  \ARIsTEO extends) to specify properties that involve variables that can assume values within a finite set of values.

The results from \Cref{table:piecewise} show that the falsification rates for 
AT1, AT2, AT51, AT52, AT53, AT6abc,
WT1, WT2, WT3, WT4, CC1, CC2, CC3, CC5, F16
are the same for all the three years of the competition.
The mean (\Savg) and median (\Smed) number of simulations (rounded down) over successful trials are also consistent for these benchmarks (small fluctuations are caused by the non-determinism of the underlying search-based algorithm).
For AT54, the property was not falsified in 2022 while it was falsified in $3$ cases in 2021. 
However, 
unlike 2021 that require the experiments to be executed $50$ times, in 2022 experiments were only executed $10$ times.
For AT6a, AT6b, AT6c, and CCX small fluctuations in the falsification rate, mean (\Savg) and median (\Smed) number of simulations are caused by the non-determinism of the underlying search-based algorithm which may produce different results across multiple experiments.
For NN, we noticed that the results of 2022 are significantly different from the ones obtained in 2020 and 2021.
Unlike 2022, in 2020 and 2021,  there was a mistake in the specification of the requirement: there was a typo in one of the temporal operators used to specify the requirement ($\square_{[1,18]}$ instead of $\square_{[1,37]}$). 
For CC4 and SC, the results of 2020 (the first year in which \ARIsTEO participated in the competition) are significantly different from the ones obtained in 2021 and 2022.
For CC4, in 2020, the predicate was written incorrectly: $y_{4}-y{5}\geq 8$ instead of $y_{5}-y{4}\geq 8$. The former is immediately falsified. Furthermore, the simulation time in 2020 was $30$s, which was lower than the temporal operators of the requirement (in 2021-22 it was $100$s). For SC, in 2020, \ARIsTEO used 12 control points and a simulation time of $100s$. In 2021-22, \ARIsTEO used 20 control points, and the simulation time was reduced to $35s$. 
Furthermore, for piecewise continuous input signals \ARIsTEO reports the results for constrained input signals: the piecewise continuous input signals let the participant be free to decide the shape of their input signals, and it was decided to use the one mandated by the constrained input signals.

The results from \Cref{table:constrained} show that the falsification rates for 
AT1, AT2, AT51, AT53, AT54, AFC29, NN, 
WT1, WT2, WT3, WT4, CC1, CC2, CC3,
are the same for all the three years of the competition.
The mean (\Savg) and median (\Smed) number of simulations (rounded down) over successful trials are also consistent for these benchmarks (small fluctuations are caused by the non-determinism of the underlying search-based algorithm).
For AT52, AT6a, AT6b, AT6c, AT6abc, AFC27, and CC5 small fluctuations in the falsification rate, mean (\Savg) and median (\Smed) number of simulations are caused by the non-determinism of the underlying search-base algorithm which may produce different results across multiple runs.
For AFC33, we noticed that the results of 2022 are significantly different from the ones obtained in 2020 and 2021.
AFC33 has a different input range compared to AFC27 and AFC29. 
In 2020 and 2021, the input range of AFC27 and AFC29 was erroneously considered.
In 2020, the wrong input range was considered (there was a bug in the code that was supposed to change the range). In 2021, the input range was correct, but the control points were considered in the wrong order.
For NNx, the results of 2022 are significantly different from the ones obtained in 2021 (NNx was not part of the benchmark models in 2020).
NNx has a different input range compared to NN. 
However, in 2021, the input range for NN was erroneously considered.
For CC4, the results of 2020 are significantly different from the ones obtained in 2021 and 2022.
As previously reported, the cause was a problem in the specification of one of the predicates and a different simulation time. 
For SC, the success rate for 2022 should be 0; there was a typo on the report for ARCH 2022.
In addition, for SC, in 2020, \ARIsTEO considered a different simulation time and a different number of control points.

\section{Lessons Learned}
\label{sec:lessons}
In this section, we reflect on the lessons learned from our replication studies and the creation of a taxonomy for the SA-SBT literature. Our lessons concern the methods and metrics used to evaluate SA-SBT  solutions, the scope and purpose of using surrogates in SBT, and the need for frameworks to facilitate extensive experimentation in this domain. We believe our lessons would be most relevant for software engineering researchers and practitioners working on the verification and testing of  cyber-physical and autonomous systems.

\subsection{Metrics for Evaluating SA-SBT}
The evaluation metrics used in the subject papers and the research reviewed in Section~\ref{sec:position} can be grouped into three categories: (1)~\emph{Metrics to assess the effectiveness of SA-SBT.} The effectiveness of a SA-SBT solution is evaluated by its ability to identify failures in the SUT and is quantified by the number of distinct failure scenarios it detects and the severity of each scenario, as indicated by the fitness function values.  (2)~\emph{Metrics to assess the efficiency of SA-SBT.} As discussed in Section~\ref{sec:taxonomy} under the \emph{evaluation metrics} category, the efficiency of SA-SBT techniques is typically evaluated based on the number of iterations, execution time and estimated execution time. (3)~\emph{Metrics to assess the performance SA-SBT solutions that rely on Pareto-based search algorithms.}  These include metrics such as HV, Spread and GD discussed in Section~\ref{sec:replicate}. Below, we discuss the limitations of some of these metrics when applied to SA-SBT solutions, and suggest alternatives to these metrics or ways to redefine these metrics to be applicable to SA-SBT solutions.

%number of failures. 
\textbf{Metrics for assessing effectiveness.} Papers on software testing often report the number of generated failures as a metric to evaluate effectiveness (e.g.,~\cite{Abdessalem:18,Ulhaq:22}).  However, for SA-SBT solutions, particularly when they are applied to case studies such as ADAS, it is important to report failures that are \emph{distinct} and \emph{valid}.   Failures are considered distinct when they are generated by test inputs that are significantly different. 
In the study that we replicated in Section~\ref{sec:replicate},  distinct scenarios are defined as those that differ in at least one input value~\cite{Abdessalem:18}. While this definition  might be sufficient for the AEB and AVP case studies in Section~\ref{sec:replicate}, it may fall short  in other situations where  two test inputs may differ in variables that are not related to failures. For example, we may have two pedestrians in the initial scene where one is far away from the collision. Changing the position of the far-away pedestrian  would not help reveal interesting and diverse failure situations. 
In addition, small variations in continuous input variables may not reveal diverse failures. For example, in Figure~\ref{fig:scenario_original}, slightly moving the pedestrian to the left or right in a failing test input may still lead to failures that are essentially the same. In a recent study, Zhong et. al.~\cite{4Zhong2021}  proposed an improved definition of distinctness that requires variations in a minimum number of input variables and differences between continuous variables to exceed a user-defined threshold. However, this definition relies on user-defined parameters that can vary between domains, making it difficult to replicate and generalize the evaluation results.

In addition to being distinct, failures should be valid and meaningful. For example, both AVP and AEB fail if the car speed is higher than the speed limit for urban areas (i.e., $100km/h$). To ensure the validity of failures, it is important that we select the input variable ranges such that they satisfy the expected preconditions of the SUT.  Unmet preconditions may cause failures that are not faults in the SUT, but rather caused by invalid inputs. The validity of failures, further,  depends on the way we formalize the notion of failure or the way we define test oracles. In our NSGAII-DT study, the criticality function acts as a test oracle and determines whether, or not, a scenario is a failure. For the AEB case study, the criticality function  was  defined using a conjunction of predicates over  fitness functions that are capped by some thresholds. The formalization of the fitness functions, however, had limitations which led to some invalid failures, e.g., the pedestrian hitting the car from the side~\cite{Abdessalem:16,Abdessalem:18}.  We addressed this issue, in our replication study (Section~\ref{sec:replicate}),  by redefining the fitness functions such that scenarios where the pedestrians approach the ego car from the sides are penalized and are not considered as critical. %It is worth noting that function defining failures (i.e., test oracles) may be defined independently  from the fitness function that guides the search. The former is a Boolean function labelling test scenarios as pass and fail, while the latter are quantitative functions guiding the test generation process. While the two are closely related, we may have situations where the criticality function is defined independently from the fitness functions. 

\textbf{Metrics for assessing efficiency.}
A primary research question for evaluating SA-SBT techniques is to assess whether they \emph{improve the efficiency of testing while ensuring its effectiveness}.   All the SA-SBT techniques we reviewed in Table~\ref{sec:classification} answer this research question by comparing SA-SBT with some non-surrogate SBT baseline. They use different metrics for this comparison though. Specifically, all the approaches, except for Menghi et. al.~\cite{9283957}, use the \emph{execution time} metric. That is, they execute the SA-SBT and the baseline techniques for the  same amount of time and compare the results. In all these cases, however, the  compute-intensive study subjects used for evaluation are simulators or Simulink models that take between two to ten minutes to execute. When the execution times of the study subjects are in the order of a few minutes, it is still feasible to  execute the SA-SBT and the baseline several times and draw conclusions using  statistical tests. It is, however,  important to note that, in these papers, the number of times experiments were repeated to account for randomness was rather low, i.e., between 10 to 30 times. This can negatively impact the accuracy of the statistical test results. Further, none of these papers performed any extensive meta-evaluations or hyper-parameter optimizations.

In the case of Menghi et. al.~\cite{9283957}, a single simulation of the compute-intensive subject takes 1.5 hour, and the approach of the paper which involves the use of SI requires extensive hyper-parameter optimizations. Performing all the required experiments on their compute-intensive subject could take in the order of 50 years to complete. Hence, Menghi et. al.~\cite{9283957} perform a part of their experiments on  non-compute-intensive subjects, measure the number of iterations required to detect failures and compute the \emph{estimated execution time}, i.e., 
the estimated time that performing these numbers of iterations would require for compute-intensive models. 
The testing techniques are then compared by considering the 
\emph{estimated execution time} instead of the same \emph{execution time}.

%Jiahui Wu, Paolo Arcaini, Tao Yue, Shaukat Ali, Huihui Zhang: On the preferences of quality indicators for multi-objective search algorithms in search-based software engineering. Empir. Softw. Eng. 27(6): 144 (2022)
%A recent study discusses the relationship between different Pareto-based search algorithms and the quality indicators and investigate the relationship between different~\cite{WuAYAZ22}. 

%Metrics for evaluating multi-objective SA-SBT for 

\textbf{Metrics for assessing the performance of Pareto-based algorithms.} The quality indicators typically determine how fast the outputs of a Pareto-based algorithm converge towards an ideal or true Pareto front. These indicators may not directly relate to the main objective of a test generation algorithm though. Hence, there needs to be a justification as to why a quality indicator is used to assess a test generation algorithm. In our original study we selected three indicators, HV, Spread and GD, based on their widespread use and their collective ability to evaluate complementary aspects of convergence, spread, and uniformit~\cite{WuAYAZ22}.  As a result of our replication, we have reached at two observations related to the use of quality indicators for assessing Pareto-based test generation algorithms: (1)~GD and Spread metrics are most meaningful for comparing population sets with equal sizes. Spread, while being normalised to the size of a population set, provides different values for two sets of different sizes, in which individuals are equally separated, which might lead to wrong interpretation of the spread for comparing populations of the same algorithm generated over subsequent runs. Even though GD does not have exactly the same problem, introduction of another individual to a population set of a small size might heavily worsen the metric value, and therefore, might also lead to incorrect interpretation of convergence over a run. We note that in contrast to classical Pareto optimization algorithms (e.g. NSGAII), NSGAII-DT allows for a flexible population size during the search. Relaxing the population size in search algorithms employed for testing is common not only for SA-SBT techniques but also in the many-objective search algorithms~\cite{AbdessalemPNBS18,PanichellaKT15}. The flexibility in the population size allows for more targeted exploration guided by surrogates, or to maintain a subset of the solutions on an optimal Pareto that are useful for the testing problem at hand. However, flexible population sizes can also affect metrics such as Spread and GD leading to potential misinterpretation of the results. (2)~Among the three quality indicators used in our original study, the results related to HV in the replication study are the most faithful to those obtained from the original study. This might be partly due to the issue that Spread and GD are not suitable metrics for comparing Pareto-based algorithms with varying populations while HV is not dependent on the population size. Overall, in our experience,  HV showed as a better metric for assessing our proposed  Pareto-based SA-SBT test generation algorithms.

\subsection{Surrogates for fault localization and repair}
The main objective of existing SA-SBT approaches is to generate  test cases that can effectively reveal failures. Empirical evidence from our subject papers~\cite{Abdessalem:16,Abdessalem:18,Matinnejad:14,9283957}, the research reviewed in Section~\ref{sec:position} and our two replication studies  support the idea that the incorporation of surrogates into SBT significantly enhances the effectiveness and efficiency of automated testing in revealing failures.  
However, it is important to ensure that the techniques do not generate spurious failures, e.g., failures caused by invalid inputs and not due to  system faults or failures that represent unrealistic situations. In our paper introducing NSGAII-DT, we utilized surrogate models to identify input conditions that led to failures, which we then validated against domain knowledge~\cite{Abdessalem:18}. These conditions are valuable because they can help explain why failures occurred, which in turn can aid in root-cause analysis, fault localization, and bug fixing. However, we did not leverage these conditions to distinguish between valid and spurious failures, nor did we use them to identify root causes of failures. To the best of our knowledge, no prior research has explored the potential of surrogate models in explaining failure conditions. We believe that these models can serve as effective tools for identifying spurious failures, as well as developing fault localization and repair strategies.

%Thus, it is crucial to design reliable and sound testing techniques that avoid the generation of spurious failures. 
%Reporting these conditions may help assess if the test case generation frameworks lead to the generation of non-spurious failures. 

\subsection{Frameworks for large-scale experimentation}
% Replications with Simulink is more mature than with ADAS. 
%Benchmarking and competitions help developing more trustworthy replication results. 
Further advancements in the area of SA-SBT require  systematic evaluation frameworks that promote accuracy, transparency and reproducibility. Our first lessons-learned highlighted the challenge of defining proper evaluation metrics, but there is also the additional obstacle of limited platforms and benchmarks available for large-scale experimentation with SA-SBT methods. As discussed in our taxonomy (Section~\ref{sec:taxonomy}), SA-SBT methods may be applied in the MiL, SiL and HiL contexts, but the main focus of the research so far has been on MiL testing. The majority of MiL testing has been performed on case studies based on simulation-based testing of autonomous driving systems or Simulink models. A variety of simulators and ADAS components have been used in the literature for simulation-based testing of autonomous driving systems. The example simulators include: PreScan~\cite{prescan}, Pro-SiVIC~\cite{belbachir_simulation-driven_2012}, CARLA~\cite{dosovitskiy2017carla}, and BeamNG~\cite{GambiMF19}, and the ADAS components include both industrial and proprietary ADAS and open source DNN-enabled components. A framework is needed to enable experimentation and evaluation of testing algorithms using different simulators and ADAS alternatives, as most current test generation algorithms are evaluated on a small subset, or even a single, simulator and ADAS.
 Our replication in Section~\ref{sec:replicate} concerned reproducing results from an original study using the same simulator but two different ADAS. As shown there, a number of different factors need to be considered when we apply a heuristic algorithm to two different ADAS even when the simulator is the same. Other replication attempts in the context of ADAS testing include the work of Borg et. al.~\cite{BorgANJS21} on replicating the results of a search-based testing algorithm on two different simulators (i.e., ProSiVic and PreScan) for the same ADAS component, and the work of Stocco et. al.~\cite{Stocco_2022} on generalizing the testing results obtained using an ADAS simulator to a physical platform. Both studies observe  notable discrepancies between the testing results obtained from the different simulators and platforms. In particular, the latter work which is a rare example of transferring the testing result from MiL to HiL shows some major reality gaps between the virtual and physical world.  A notable effort in evaluating and comparing MiL, SiL and HiL test setups is the recent work by Mandrioli et. al.~\cite{9502570}. This work which is performed in the context of drones demonstrates that the results obtained from MiL, SiL and HiL levels have notable differences, and in fact, these levels have complementary capabilities in revealing faults for CPS.

\subsection{International Competitions}
International competitions are a successful instrument to support experiment replication effectively.
While participating in competitions requires significant effort, these competitions offer several benefits. 
We learned many lessons over our three-year participation in the ARCH competition that we summarize below. 

\emph{Forces tools updates}.
Authors need to constantly update their tools to ensure specific formatting of their inputs and outputs to adhere to the rules of the competition. 
From our experience, we learned that this activity improved our tool. 
 For example, participating in the competition in  2021 enabled us to extend the implementation of \ARIsTEO to support property specifications that use locations.
 This activity extended the support of \ARIsTEO to benchmarks AT51, AT52, AT53, and AT54. 

\emph{Supports experiment replication and verification of previous results}.
 Replicating previous experiments is not always a trivial and straightforward activity. 
Replicating the experiments for AT51, AT52, AT54, and AT54 was not trivial. 
\ARIsTEO is a plugin for \staliro.
 \staliro requires to specify  these requirements using ``control locations'' (Pred(i).loc), i.e., specific constructs used to establish fitness metrics of hybrid systems.
Despite \staliro being widely documented and the explanations being detailed and exhaustive, the documentation of this construct should be improved.
Specifically, to analyze these requirements, we had to install MatlabBGL~\cite{MatlabBGL}, a library containing a set of algorithms to work with graphs, and recompile the C$++$ files that are used to compute the fitness metric using an old version (1\_36\_0~\cite{boost}) of Boost, a portable C$++$ and GCC library. 
However, this needed to be more precisely detailed since we had to reverse engineer the version of the library to be used: we could not use newer versions of Boost since they are not retro-compatible for this case.
\ARIsTEO participation was managed by different students across the years. 
We spent significant time in 2021 adding support for these requirements by extending the implementation of \ARIsTEO and understanding how to install the different libraries. 
However, we did not precisely document our activity, and we had to perform the same reverse engineering process in 2022.
We now have a detailed description of the libraries that need to be installed to support these requirements, add it to our documentation, and contact the developers of \staliro to extend their documentation. 
Finally, as detailed in \Cref{sec:simulink}, participating in consecutive years of the competition enables spotting configuration errors and improving the reliability of the published results.

\emph{Discovering new results over time.}
The set of benchmark models change over time: the ARCH competition is organized as part of the International Workshop on Applied Verification of Continuous and Hybrid Systems, which contains an explicit call for new benchmark models. 
For example, in 2022, the organizers removed the WT model from the set of benchmark models considered in the competition, and the pacemaker model~\cite{ARCH22:Two_Simulink_Models_with} was proposed at the workshop and is likely to be added to the set of benchmark models to be considered in future editions of the competition.
The variation of the benchmark models enables a continuous comparison and assessment of the tools on models from different domains and with various characteristics (e.g., number of blocks).

\emph{Supports networking and research collaborations}. 
The ARCH competition is a friendly competition with the primary purpose of keeping researchers and practitioners updated with the latest advancements in the field.
It is designed to support and foster networking and research collaborations.
This decision ensures that participants are offered a friendly and cordial environment.
We learned that this offered a great opportunity (especially for students) for networking, getting new research ideas, and establishing research collaborations.

\vspace{0.5cm}
Finally, we  learned that organizing and managing the competition is a significant undertaking. 
The coordinator of the Falsification Category of the ARCH Competition (Gidon Ernst) did a significant job across the three years in which we participated to ensure the competition's success. Organizing the meetings to define the rules of the competition, collecting and analyzing the results, leading the writing of the competition report, and reporting the results at the workshop are a significant service to the community that should be acknowledged and appropriately rewarded. 

\vspace{0.5cm}
Our taxonomy can support the organizers of the ARCH competition by supporting activities such as adding new benchmarks. For example, the organizers may extend the benchmarks to include models with more complexity and longer execution time. 
Further, the metrics we discussed earlier in this section can be used as part of the benchmark and competition since, currently, the number of iterations is the only metric used for the tool competition.

% One promising way to achieve this goal is through benchmarking~\cite{SimEH03}. The latter benchmark has been used for a tool competition on falsification techniques which are instances of SBT focused on testing CPS models against their temporal logic specifications.  One of the SA-SBT techniques we reviewed in this paper (i.e.,~\cite{9283957}) participated in the last two rounds of the ARCH competition~\cite{DBLP:conf/arch/ErnstABCDFFG0KM21,DBLP:conf/arch/ErnstABDFFMMPPY20}.

%\section{Emerging ideas and current vision}
%\label{sec:emerge}
%\input{Sections/emerge}

\section{Conclusions}
\label{sec:con}
This paper reflects  on four papers published between 2014 and 2021~\cite{Abdessalem:16,Abdessalem:18,9283957,Matinnejad:14} that propose surrogate-assisted search-based testing (SA-SBT) 
techniques for autonomous systems. We developed a taxonomy based on  our synthesis of different SA-SBT approaches. Our taxonomy identifies the main dimensions of SA-SBT solutions and we demonstrate how different SA-SBT solutions can be positioned along these dimensions. We report on two replication studies using an industrial advanced driver assistance system (ADAS) and a benchmark of Simulink models. The results of our taxonomy and replication studies highlight the need for improvement in the metrics used to evaluate SA-SBT, the potential for using surrogates in fault localization and repair, and the importance of benchmarking, international competitions and creating frameworks for large-scale experiments involving SA-SBT techniques.

\section*{Acknowledgements}
We acknowledge Gidon Ernst for leading the Falsification Category of the ARCH Competition in 2020, 2021, and 2022. We also acknowledge Khouloud Gaaloul that supported the participation of \ARIsTEO in the 2020 and 2021 editions of the ARCH competition.

We acknowledge the support of the Natural Sciences and Engineering Research Council of Canada (NSERC) [funding reference numbers RGPIN-2022-04622,DGECR-2022-0040]. 
This research paper has further received funding from the European Union’s Horizon 2020 research and innovation programme under grant agreement No 956123. % TODO add european embleme
\bibliographystyle{elsarticle-harv}
\bibliography{bibliography}

\begin{thebibliography}{55}
\expandafter\ifx\csname natexlab\endcsname\relax\def\natexlab#1{#1}\fi
\providecommand{\url}[1]{\texttt{#1}}
\providecommand{\href}[2]{#2}
\providecommand{\path}[1]{#1}
\providecommand{\DOIprefix}{doi:}
\providecommand{\ArXivprefix}{arXiv:}
\providecommand{\URLprefix}{URL: }
\providecommand{\Pubmedprefix}{pmid:}
\providecommand{\doi}[1]{\href{http://dx.doi.org/#1}{\path{#1}}}
\providecommand{\Pubmed}[1]{\href{pmid:#1}{\path{#1}}}
\providecommand{\bibinfo}[2]{#2}
\ifx\xfnm\relax \def\xfnm[#1]{\unskip,\space#1}\fi
%Type = Misc
\bibitem[{boo(2022)}]{boost}
, \bibinfo{year}{2022}.
\newblock \bibinfo{title}{boost}.
\newblock
  \bibinfo{howpublished}{\url{https://sourceforge.net/projects/boost/files/boost/1.36.0/}}.
\newblock \bibinfo{note}{Accessed: 2022-11-21}.
%Type = Misc
\bibitem[{NN(2022)}]{NN}
, \bibinfo{year}{2022}.
\newblock \bibinfo{title}{Design narma-l2 neural controller in simulink}.
\newblock
  \bibinfo{howpublished}{\url{https://www.mathworks.com/help/deeplearning/ug/design-narma-l2-neural-controller-in-simulink.html}}.
\newblock \bibinfo{note}{Accessed: 2022-11-21}.
%Type = Misc
\bibitem[{Mat(2022)}]{MatlabBGL}
, \bibinfo{year}{2022}.
\newblock \bibinfo{title}{Matlabbgl}.
\newblock
  \bibinfo{howpublished}{\url{https://www.mathworks.com/matlabcentral/fileexchange/10922-matlabbgl}}.
\newblock \bibinfo{note}{Accessed: 2022-11-21}.
%Type = Misc
\bibitem[{res(2023)}]{resultsReplication1}
, \bibinfo{year}{2023}.
\newblock \bibinfo{title}{Results replication {NSGAII-DT}}.
\newblock \URLprefix \url{https://github.com/Leviathan321/reflection_study}.
%Type = Misc
\bibitem[{pre(2023)}]{prescan}
, \bibinfo{year}{2023}.
\newblock \bibinfo{title}{{Simcenter Prescan}}.
\newblock \URLprefix
  \url{https://www.plm.automation.siemens.com/global/de/products/simcenter/prescan.html}.
  \bibinfo{note}{accessed: 2023-02-02}.
%Type = Inproceedings
\bibitem[{Abdessalem et~al.(2016)Abdessalem, Nejati, Briand and
  Stifter}]{Abdessalem:16}
\bibinfo{author}{Abdessalem, R.B.}, \bibinfo{author}{Nejati, S.},
  \bibinfo{author}{Briand, L.C.}, \bibinfo{author}{Stifter, T.},
  \bibinfo{year}{2016}.
\newblock \bibinfo{title}{Testing advanced driver assistance systems using
  multi-objective search and neural networks}, in:
  \bibinfo{booktitle}{International Conference on Automated Software
  Engineering ({ASE})}, \bibinfo{publisher}{{IEEE/ACM}}. pp.
  \bibinfo{pages}{63--74}.
%Type = Inproceedings
\bibitem[{Abdessalem et~al.(2018a)Abdessalem, Nejati, Briand and
  Stifter}]{Abdessalem:18}
\bibinfo{author}{Abdessalem, R.B.}, \bibinfo{author}{Nejati, S.},
  \bibinfo{author}{Briand, L.C.}, \bibinfo{author}{Stifter, T.},
  \bibinfo{year}{2018}a.
\newblock \bibinfo{title}{Testing vision-based control systems using learnable
  evolutionary algorithms}, in: \bibinfo{editor}{Chaudron, M.},
  \bibinfo{editor}{Crnkovic, I.}, \bibinfo{editor}{Chechik, M.},
  \bibinfo{editor}{Harman, M.} (Eds.), \bibinfo{booktitle}{International
  Conference on Software Engineering, ({ICSE})}, \bibinfo{publisher}{{ACM}}.
  pp. \bibinfo{pages}{1016--1026}.
%Type = Inproceedings
\bibitem[{Abdessalem et~al.(2018b)Abdessalem, Panichella, Nejati, Briand and
  Stifter}]{AbdessalemPNBS18}
\bibinfo{author}{Abdessalem, R.B.}, \bibinfo{author}{Panichella, A.},
  \bibinfo{author}{Nejati, S.}, \bibinfo{author}{Briand, L.C.},
  \bibinfo{author}{Stifter, T.}, \bibinfo{year}{2018}b.
\newblock \bibinfo{title}{Testing autonomous cars for feature interaction
  failures using many-objective search}, in: \bibinfo{editor}{Huchard, M.},
  \bibinfo{editor}{K{\"{a}}stner, C.}, \bibinfo{editor}{Fraser, G.} (Eds.),
  \bibinfo{booktitle}{International Conference on Automated Software
  Engineering ({ASE})}, \bibinfo{publisher}{{ACM/IEEE}}. pp.
  \bibinfo{pages}{143--154}.
%Type = Article
\bibitem[{AMS and Electromagnetism(1989)}]{ams1989system}
\bibinfo{author}{AMS, M.M.}, \bibinfo{author}{Electromagnetism, P.},
  \bibinfo{year}{1989}.
\newblock \bibinfo{title}{System identification} .
%Type = Inproceedings
\bibitem[{Annpureddy et~al.(2011)Annpureddy, Liu, Fainekos and
  Sankaranarayanan}]{S-Taliro}
\bibinfo{author}{Annpureddy, Y.}, \bibinfo{author}{Liu, C.},
  \bibinfo{author}{Fainekos, G.}, \bibinfo{author}{Sankaranarayanan, S.},
  \bibinfo{year}{2011}.
\newblock \bibinfo{title}{S-taliro: A tool for temporal logic falsification for
  hybrid systems}, in: \bibinfo{booktitle}{Tools and Algorithms for the
  Construction and Analysis of Systems}, \bibinfo{publisher}{Springer}. pp.
  \bibinfo{pages}{254--257}.
%Type = Misc
\bibitem[{ARCH()}]{ARCHWEBSITE}
ARCH, \bibinfo{year}{2022 [Online]}.
\newblock \bibinfo{title}{{International Competition on Verifying Continuous
  and Hybrid Systems}}.
\newblock \URLprefix \url{https://cps-vo.org/group/ARCH/FriendlyCompetition}.
%Type = Inproceedings
\bibitem[{Arrieta et~al.(2017)Arrieta, Wang, Markiegi, Sagardui and
  Etxeberria}]{3Arrieta2017}
\bibinfo{author}{Arrieta, A.}, \bibinfo{author}{Wang, S.},
  \bibinfo{author}{Markiegi, U.}, \bibinfo{author}{Sagardui, G.},
  \bibinfo{author}{Etxeberria, L.}, \bibinfo{year}{2017}.
\newblock \bibinfo{title}{Search-based test case generation for cyber-physical
  systems}, in: \bibinfo{booktitle}{Congress on Evolutionary Computation
  ({CEC})}, \bibinfo{publisher}{{IEEE}}. pp. \bibinfo{pages}{688--697}.
%Type = Inproceedings
\bibitem[{Ayesh et~al.(2022)Ayesh, Mehan, Dhanraj, El-Rahwan, Opalka, Fan,
  Hamilton, Jacob, Sundarrajan, Widjaja and
  Menghi}]{ARCH22:Two_Simulink_Models_with}
\bibinfo{author}{Ayesh, M.}, \bibinfo{author}{Mehan, N.},
  \bibinfo{author}{Dhanraj, E.}, \bibinfo{author}{El-Rahwan, A.},
  \bibinfo{author}{Opalka, S.E.}, \bibinfo{author}{Fan, T.},
  \bibinfo{author}{Hamilton, A.}, \bibinfo{author}{Jacob, A.M.},
  \bibinfo{author}{Sundarrajan, R.A.}, \bibinfo{author}{Widjaja, B.},
  \bibinfo{author}{Menghi, C.}, \bibinfo{year}{2022}.
\newblock \bibinfo{title}{Two simulink models with requirements for a simple
  controller of a pacemaker device}, in: \bibinfo{booktitle}{International
  Workshop on Applied Verification of Continuous and Hybrid Systems (ARCH22)},
  \bibinfo{publisher}{EasyChair}. pp. \bibinfo{pages}{18--25}.
%Type = Inproceedings
\bibitem[{Beglerovic et~al.(2017)Beglerovic, Stolz and Horn}]{1Beglerovic2017}
\bibinfo{author}{Beglerovic, H.}, \bibinfo{author}{Stolz, M.},
  \bibinfo{author}{Horn, M.}, \bibinfo{year}{2017}.
\newblock \bibinfo{title}{Testing of autonomous vehicles using surrogate models
  and stochastic optimization}, in: \bibinfo{booktitle}{International
  Conference on Intelligent Transportation Systems ({ITSC})},
  \bibinfo{publisher}{{IEEE}}. pp. \bibinfo{pages}{1--6}.
%Type = Article
\bibitem[{Belbachir et~al.(2012)Belbachir, Smal, Blosseville and
  Gruyer}]{belbachir_simulation-driven_2012}
\bibinfo{author}{Belbachir, A.}, \bibinfo{author}{Smal, J.C.},
  \bibinfo{author}{Blosseville, J.M.}, \bibinfo{author}{Gruyer, D.},
  \bibinfo{year}{2012}.
\newblock \bibinfo{title}{Simulation-{Driven} {Validation} of {Advanced}
  {Driving}-{Assistance} {Systems}}.
\newblock \bibinfo{journal}{Procedia - Social and Behavioral Sciences}
  \bibinfo{volume}{48}, \bibinfo{pages}{1205--1214}.
\newblock \DOIprefix\doi{10.1016/j.sbspro.2012.06.1096}.
%Type = Book
\bibitem[{Bittanti(2019)}]{bittanti2019model}
\bibinfo{author}{Bittanti, S.}, \bibinfo{year}{2019}.
\newblock \bibinfo{title}{Model identification and data analysis}.
\newblock \bibinfo{publisher}{John Wiley \& Sons}.
%Type = Inproceedings
\bibitem[{Borg et~al.(2021)Borg, Abdessalem, Nejati, Jegeden and
  Shin}]{BorgANJS21}
\bibinfo{author}{Borg, M.}, \bibinfo{author}{Abdessalem, R.B.},
  \bibinfo{author}{Nejati, S.}, \bibinfo{author}{Jegeden, F.},
  \bibinfo{author}{Shin, D.}, \bibinfo{year}{2021}.
\newblock \bibinfo{title}{Digital twins are not monozygotic - cross-replicating
  {ADAS} testing in two industry-grade automotive simulators}, in:
  \bibinfo{booktitle}{Conference on Software Testing, Verification and
  Validation ({ICST})}, \bibinfo{publisher}{{IEEE}}. pp.
  \bibinfo{pages}{383--393}.
%Type = Article
\bibitem[{Chen et~al.(2020)Chen, Li and Yao}]{TaoParetoQI20}
\bibinfo{author}{Chen, T.}, \bibinfo{author}{Li, M.}, \bibinfo{author}{Yao,
  X.}, \bibinfo{year}{2020}.
\newblock \bibinfo{title}{How to evaluate solutions in pareto-based
  search-based software engineering? {A} critical review and methodological
  guidance}.
\newblock \bibinfo{journal}{CoRR} \bibinfo{volume}{abs/2002.09040}.
\newblock \URLprefix \url{https://arxiv.org/abs/2002.09040},
  \href{http://arxiv.org/abs/2002.09040}{{\tt arXiv:2002.09040}}.
%Type = Article
\bibitem[{Clarke et~al.(2001)Clarke, Biere, Raimi and Zhu}]{Clarke:01}
\bibinfo{author}{Clarke, E.M.}, \bibinfo{author}{Biere, A.},
  \bibinfo{author}{Raimi, R.}, \bibinfo{author}{Zhu, Y.}, \bibinfo{year}{2001}.
\newblock \bibinfo{title}{Bounded model checking using satisfiability solving}.
\newblock \bibinfo{journal}{Formal Methods in System Design}
  \bibinfo{volume}{19}, \bibinfo{pages}{7--34}.
%Type = Article
\bibitem[{Deb et~al.(2002)Deb, Pratap, Agarwal and Meyarivan}]{DebNSGA2_02}
\bibinfo{author}{Deb, K.}, \bibinfo{author}{Pratap, A.},
  \bibinfo{author}{Agarwal, S.}, \bibinfo{author}{Meyarivan, T.},
  \bibinfo{year}{2002}.
\newblock \bibinfo{title}{A fast and elitist multiobjective genetic algorithm:
  {NSGA-II}}.
\newblock \bibinfo{journal}{IEEE Transactions on Evolutionary Computation}
  \bibinfo{volume}{6}, \bibinfo{pages}{182--197}.
\newblock \DOIprefix\doi{10.1109/4235.996017}.
%Type = Article
\bibitem[{Dosovitskiy et~al.(2017)Dosovitskiy, Ros, Codevilla, Lopez and
  Koltun}]{dosovitskiy2017carla}
\bibinfo{author}{Dosovitskiy, A.}, \bibinfo{author}{Ros, G.},
  \bibinfo{author}{Codevilla, F.}, \bibinfo{author}{Lopez, A.},
  \bibinfo{author}{Koltun, V.}, \bibinfo{year}{2017}.
\newblock \bibinfo{title}{Carla: An open urban driving simulator}.
\newblock \bibinfo{journal}{arXiv preprint arXiv:1711.03938} .
%Type = Article
\bibitem[{Dyb{\aa} et~al.(2014)Dyb{\aa}, Maiden and
  Glass}]{DBLP:journals/software/DybaMG14}
\bibinfo{author}{Dyb{\aa}, T.}, \bibinfo{author}{Maiden, N.A.M.},
  \bibinfo{author}{Glass, R.}, \bibinfo{year}{2014}.
\newblock \bibinfo{title}{The reflective software engineer: Reflective
  practice}.
\newblock \bibinfo{journal}{{IEEE} Softw.} \bibinfo{volume}{31},
  \bibinfo{pages}{32--36}.
%Type = Inproceedings
\bibitem[{Ernst et~al.(2020)Ernst, Arcaini, Bennani, Donz{\'{e}}, Fainekos,
  Frehse, Mathesen, Menghi, Pedrielli, Pouzet, Yaghoubi, Yamagata and
  Zhang}]{DBLP:conf/arch/ErnstABDFFMMPPY20}
\bibinfo{author}{Ernst, G.}, \bibinfo{author}{Arcaini, P.},
  \bibinfo{author}{Bennani, I.}, \bibinfo{author}{Donz{\'{e}}, A.},
  \bibinfo{author}{Fainekos, G.}, \bibinfo{author}{Frehse, G.},
  \bibinfo{author}{Mathesen, L.}, \bibinfo{author}{Menghi, C.},
  \bibinfo{author}{Pedrielli, G.}, \bibinfo{author}{Pouzet, M.},
  \bibinfo{author}{Yaghoubi, S.}, \bibinfo{author}{Yamagata, Y.},
  \bibinfo{author}{Zhang, Z.}, \bibinfo{year}{2020}.
\newblock \bibinfo{title}{{ARCH-COMP} 2020 category report: Falsification}, in:
  \bibinfo{booktitle}{{International Workshop on Applied Verification of
  Continuous And Hybrid Systems}}, \bibinfo{publisher}{EasyChair}. pp.
  \bibinfo{pages}{140--152}.
%Type = Inproceedings
\bibitem[{Ernst et~al.(2022)Ernst, Arcaini, Fainekos, Formica, Inoue, Khandait,
  Mahboob, Menghi, Pedrielli, Waga, Yamagata and
  Zhang}]{ARCH22:ARCH_COMP_2022_Category_Report}
\bibinfo{author}{Ernst, G.}, \bibinfo{author}{Arcaini, P.},
  \bibinfo{author}{Fainekos, G.}, \bibinfo{author}{Formica, F.},
  \bibinfo{author}{Inoue, J.}, \bibinfo{author}{Khandait, T.},
  \bibinfo{author}{Mahboob, M.M.}, \bibinfo{author}{Menghi, C.},
  \bibinfo{author}{Pedrielli, G.}, \bibinfo{author}{Waga, M.},
  \bibinfo{author}{Yamagata, Y.}, \bibinfo{author}{Zhang, Z.},
  \bibinfo{year}{2022}.
\newblock \bibinfo{title}{Arch-comp 2022 category report: Falsification with
  ubounded resources}, in: \bibinfo{booktitle}{International Workshop on
  Applied Verification of Continuous and Hybrid Systems (ARCH22)},
  \bibinfo{publisher}{EasyChair}. pp. \bibinfo{pages}{204--221}.
\newblock \DOIprefix\doi{10.29007/fhnk}.
%Type = Article
\bibitem[{Ernst et~al.(2018)Ernst, Sedwards, Zhang and
  Hasuo}]{ErnstSZH2018-FalStar}
\bibinfo{author}{Ernst, G.}, \bibinfo{author}{Sedwards, S.},
  \bibinfo{author}{Zhang, Z.}, \bibinfo{author}{Hasuo, I.},
  \bibinfo{year}{2018}.
\newblock \bibinfo{title}{Fast falsification of hybrid systems using
  probabilistically adaptive input}.
\newblock \bibinfo{journal}{arXiv preprint arXiv:1812.04159} .
%Type = Inproceedings
\bibitem[{G.~Ernst et~al.(2021)G.~Ernst, Donz{\'{e}}, Fainekos, Frehse,
  Gaaloul, Inoue, Khandait, Mathesen, Menghi, Pedrielli, Pouzet, Waga,
  Yaghoubi, Yamagata and Zhang}]{DBLP:conf/arch/ErnstABCDFFG0KM21}
\bibinfo{author}{G.~Ernst, P.~Arcaini, I.B.A.C.}, \bibinfo{author}{Donz{\'{e}},
  A.}, \bibinfo{author}{Fainekos, G.}, \bibinfo{author}{Frehse, G.},
  \bibinfo{author}{Gaaloul, K.}, \bibinfo{author}{Inoue, J.},
  \bibinfo{author}{Khandait, T.}, \bibinfo{author}{Mathesen, L.},
  \bibinfo{author}{Menghi, C.}, \bibinfo{author}{Pedrielli, G.},
  \bibinfo{author}{Pouzet, M.}, \bibinfo{author}{Waga, M.},
  \bibinfo{author}{Yaghoubi, S.}, \bibinfo{author}{Yamagata, Y.},
  \bibinfo{author}{Zhang, Z.}, \bibinfo{year}{2021}.
\newblock \bibinfo{title}{{ARCH-COMP} category report: Falsification with
  validation of results}, in: \bibinfo{booktitle}{Workshop on Applied
  Verification of Continuous and Hybrid Systems},
  \bibinfo{publisher}{EasyChair}. pp. \bibinfo{pages}{133--152}.
%Type = Inproceedings
\bibitem[{Gambi et~al.(2019)Gambi, M{\"{u}}ller and Fraser}]{GambiMF19}
\bibinfo{author}{Gambi, A.}, \bibinfo{author}{M{\"{u}}ller, M.},
  \bibinfo{author}{Fraser, G.}, \bibinfo{year}{2019}.
\newblock \bibinfo{title}{{AsFault}: testing self-driving car software using
  search-based procedural content generation}, in: \bibinfo{editor}{Atlee,
  J.M.}, \bibinfo{editor}{Bultan, T.}, \bibinfo{editor}{Whittle, J.} (Eds.),
  \bibinfo{booktitle}{International Conference on Software Engineering:
  Companion Proceedings (ICSE)}, \bibinfo{publisher}{{IEEE} / {ACM}}. pp.
  \bibinfo{pages}{27--30}.
%Type = Inproceedings
\bibitem[{Haq et~al.(2022)Haq, Shin and Briand}]{Ulhaq:22}
\bibinfo{author}{Haq, F.U.}, \bibinfo{author}{Shin, D.},
  \bibinfo{author}{Briand, L.}, \bibinfo{year}{2022}.
\newblock \bibinfo{title}{Efficient online testing for dnn-enabled systems
  using surrogate-assisted and many-objective optimization}, in:
  \bibinfo{booktitle}{International Conference on Software Engineering (ICSE
  2022)}, pp. \bibinfo{pages}{811--822}.
%Type = Inproceedings
\bibitem[{Heidlauf et~al.(2018)Heidlauf, Collins, Bolender and
  Bak}]{ARCH18:Verification_Challenges_in_F_16}
\bibinfo{author}{Heidlauf, P.}, \bibinfo{author}{Collins, A.},
  \bibinfo{author}{Bolender, M.}, \bibinfo{author}{Bak, S.},
  \bibinfo{year}{2018}.
\newblock \bibinfo{title}{Verification challenges in {F-16} ground collision
  avoidance and other automated maneuvers}, in: \bibinfo{editor}{Frehse, G.}
  (Ed.), \bibinfo{booktitle}{{ARCH18}. 5th International Workshop on Applied
  Verification of Continuous and Hybrid Systems},
  \bibinfo{publisher}{EasyChair}. pp. \bibinfo{pages}{208--217}.
\newblock \DOIprefix\doi{10.29007/91x9}.
%Type = Inproceedings
\bibitem[{Henard et~al.(2013)Henard, Papadakis, Perrouin, Klein and
  Traon}]{henard:13}
\bibinfo{author}{Henard, C.}, \bibinfo{author}{Papadakis, M.},
  \bibinfo{author}{Perrouin, G.}, \bibinfo{author}{Klein, J.},
  \bibinfo{author}{Traon, Y.L.}, \bibinfo{year}{2013}.
\newblock \bibinfo{title}{{PLEDGE:} a product line editor and test generation
  tool}, in: \bibinfo{booktitle}{International Software Product Line Conference
  co-located workshops ({SPLC'13})}, \bibinfo{publisher}{ACM}. pp.
  \bibinfo{pages}{126--129}.
%Type = Inproceedings
\bibitem[{Hoxha et~al.(2015)Hoxha, Abbas and Fainekos}]{ARCH14}
\bibinfo{author}{Hoxha, B.}, \bibinfo{author}{Abbas, H.},
  \bibinfo{author}{Fainekos, G.}, \bibinfo{year}{2015}.
\newblock \bibinfo{title}{Benchmarks for temporal logic requirements for
  automotive systems}, in: \bibinfo{editor}{Frehse, G.},
  \bibinfo{editor}{Althoff, M.} (Eds.), \bibinfo{booktitle}{{ARCH14-15}. 1st
  and 2nd International Workshop on Applied veRification for Continuous and
  Hybrid Systems}, \bibinfo{publisher}{EasyChair}. pp. \bibinfo{pages}{25--30}.
\newblock \DOIprefix\doi{10.29007/xwrs}.
%Type = Inproceedings
\bibitem[{Hu et~al.(2000)Hu, Lygeros and Sastry}]{hu2000towards}
\bibinfo{author}{Hu, J.}, \bibinfo{author}{Lygeros, J.},
  \bibinfo{author}{Sastry, S.}, \bibinfo{year}{2000}.
\newblock \bibinfo{title}{Towards a theory of stochastic hybrid systems}, in:
  \bibinfo{booktitle}{International Workshop on Hybrid Systems: Computation and
  Control}, \bibinfo{organization}{Springer}. pp. \bibinfo{pages}{160--173}.
%Type = Article
\bibitem[{Humeniuk et~al.(2021)Humeniuk, Antoniol and Khomh}]{7Humeniuk2021}
\bibinfo{author}{Humeniuk, D.}, \bibinfo{author}{Antoniol, G.},
  \bibinfo{author}{Khomh, F.}, \bibinfo{year}{2021}.
\newblock \bibinfo{title}{Data driven testing of cyber physical systems}.
\newblock \bibinfo{journal}{CoRR} \bibinfo{volume}{abs/2102.11491}.
\newblock \URLprefix \url{https://arxiv.org/abs/2102.11491},
  \href{http://arxiv.org/abs/2102.11491}{{\tt arXiv:2102.11491}}.
%Type = Article
\bibitem[{Humeniuk et~al.(2022)Humeniuk, Khomh and Antoniol}]{8Humeniuk2022}
\bibinfo{author}{Humeniuk, D.}, \bibinfo{author}{Khomh, F.},
  \bibinfo{author}{Antoniol, G.}, \bibinfo{year}{2022}.
\newblock \bibinfo{title}{A search-based framework for automatic generation of
  testing environments for cyber-physical systems}.
\newblock \bibinfo{journal}{Inf. Softw. Technol.} \bibinfo{volume}{149},
  \bibinfo{pages}{106936}.
\newblock \DOIprefix\doi{10.1016/j.infsof.2022.106936}.
%Type = Inproceedings
\bibitem[{Innes and Ramamoorthy(2022)}]{12Innes2022}
\bibinfo{author}{Innes, C.}, \bibinfo{author}{Ramamoorthy, S.},
  \bibinfo{year}{2022}.
\newblock \bibinfo{title}{Automated testing with temporal logic specifications
  for robotic controllers using adaptive experiment design}, in:
  \bibinfo{booktitle}{International Conference on Robotics and Automation
  ({ICRA})}, \bibinfo{publisher}{{IEEE}}. pp. \bibinfo{pages}{6814--6821}.
\newblock \DOIprefix\doi{10.1109/ICRA46639.2022.9811579}.
%Type = Inproceedings
\bibitem[{Jin et~al.(2014)Jin, Deshmukh, Kapinski, Ueda and
  Butts}]{JinHSCC2014}
\bibinfo{author}{Jin, X.}, \bibinfo{author}{Deshmukh, J.V.},
  \bibinfo{author}{Kapinski, J.}, \bibinfo{author}{Ueda, K.},
  \bibinfo{author}{Butts, K.}, \bibinfo{year}{2014}.
\newblock \bibinfo{title}{Powertrain control verification benchmark}, in:
  \bibinfo{booktitle}{International Conference on Hybrid Systems: Computation
  and Control}, \bibinfo{publisher}{ACM}. pp. \bibinfo{pages}{253--262}.
\newblock \DOIprefix\doi{10.1145/2562059.2562140}.
%Type = Article
\bibitem[{Jin(2011)}]{jin2011surrogate}
\bibinfo{author}{Jin, Y.}, \bibinfo{year}{2011}.
\newblock \bibinfo{title}{Surrogate-assisted evolutionary computation: Recent
  advances and future challenges}.
\newblock \bibinfo{journal}{Swarm and Evolutionary Computation}
  \bibinfo{volume}{1}, \bibinfo{pages}{61--70}.
%Type = Book
\bibitem[{Luke(2013)}]{Luke:13}
\bibinfo{author}{Luke, S.}, \bibinfo{year}{2013}.
\newblock \bibinfo{title}{Essentials of Metaheuristics}.
\newblock \bibinfo{edition}{second} ed., \bibinfo{publisher}{Lulu}.
\newblock \bibinfo{note}{Available for free at
  http://cs.gmu.edu/$\sim$sean/book/metaheuristics/}.
%Type = Misc
\bibitem[{Mandrioli et~al.()Mandrioli, Carlsson and Maggio}]{9502570}
\bibinfo{author}{Mandrioli, C.}, \bibinfo{author}{Carlsson, M.N.},
  \bibinfo{author}{Maggio, M.}, .
\newblock \bibinfo{title}{Testing abstractions for cyber-physical control
  systems}.
\newblock \bibinfo{note}{Submitted for publication}.
%Type = Inproceedings
\bibitem[{Matinnejad et~al.(2014)Matinnejad, Nejati, Briand and
  Bruckmann}]{Matinnejad:14}
\bibinfo{author}{Matinnejad, R.}, \bibinfo{author}{Nejati, S.},
  \bibinfo{author}{Briand, L.C.}, \bibinfo{author}{Bruckmann, T.},
  \bibinfo{year}{2014}.
\newblock \bibinfo{title}{Mil testing of highly configurable continuous
  controllers: scalable search using surrogate models}, in:
  \bibinfo{editor}{Crnkovic, I.}, \bibinfo{editor}{Chechik, M.},
  \bibinfo{editor}{Gr{\"{u}}nbacher, P.} (Eds.),
  \bibinfo{booktitle}{International Conference on Automated Software
  Engineering ({ASE})}, \bibinfo{publisher}{ACM/IEEE}. pp.
  \bibinfo{pages}{163--174}.
%Type = Article
\bibitem[{McKay et~al.(1979)McKay, Beckman and Conover}]{McKay79LHS}
\bibinfo{author}{McKay, M.D.}, \bibinfo{author}{Beckman, R.J.},
  \bibinfo{author}{Conover, W.J.}, \bibinfo{year}{1979}.
\newblock \bibinfo{title}{A comparison of three methods for selecting values of
  input variables in the analysis of output from a computer code}.
\newblock \bibinfo{journal}{Technometrics} \bibinfo{volume}{21},
  \bibinfo{pages}{239--245}.
\newblock \URLprefix \url{http://www.jstor.org/stable/1268522}.
%Type = Inproceedings
\bibitem[{Menghi et~al.(2020)Menghi, Nejati, Briand and Parache}]{9283957}
\bibinfo{author}{Menghi, C.}, \bibinfo{author}{Nejati, S.},
  \bibinfo{author}{Briand, L.}, \bibinfo{author}{Parache, Y.I.},
  \bibinfo{year}{2020}.
\newblock \bibinfo{title}{Approximation-refinement testing of compute-intensive
  cyber-physical models: An approach based on system identification}, in:
  \bibinfo{booktitle}{IEEE/ACM 42nd International Conference on Software
  Engineering (ICSE)}, pp. \bibinfo{pages}{372--384}.
%Type = Inproceedings
\bibitem[{Panichella et~al.(2015)Panichella, Kifetew and
  Tonella}]{PanichellaKT15}
\bibinfo{author}{Panichella, A.}, \bibinfo{author}{Kifetew, F.M.},
  \bibinfo{author}{Tonella, P.}, \bibinfo{year}{2015}.
\newblock \bibinfo{title}{Reformulating branch coverage as a many-objective
  optimization problem}, in: \bibinfo{booktitle}{International Conference on
  Software Testing, Verification and Validation (ICST)},
  \bibinfo{publisher}{{IEEE} Computer Society}. pp. \bibinfo{pages}{1--10}.
%Type = Article
\bibitem[{Pedrielli et~al.(2021)Pedrielli, Khandait, Chotaliya, Thibeault,
  Huang, Castillo{-}Effen and Fainekos}]{6Pedrielli2021}
\bibinfo{author}{Pedrielli, G.}, \bibinfo{author}{Khandait, T.},
  \bibinfo{author}{Chotaliya, S.}, \bibinfo{author}{Thibeault, Q.},
  \bibinfo{author}{Huang, H.}, \bibinfo{author}{Castillo{-}Effen, M.},
  \bibinfo{author}{Fainekos, G.}, \bibinfo{year}{2021}.
\newblock \bibinfo{title}{Part-x: {A} family of stochastic algorithms for
  search-based test generation with probabilistic guarantees}.
\newblock \bibinfo{journal}{CoRR} \bibinfo{volume}{abs/2110.10729}.
\newblock \URLprefix \url{https://arxiv.org/abs/2110.10729},
  \href{http://arxiv.org/abs/2110.10729}{{\tt arXiv:2110.10729}}.
%Type = Inproceedings
\bibitem[{Schuler et~al.(2017)Schuler, Adegas and
  Anta}]{ARCH16:Hybrid_Modelling_of_Wind}
\bibinfo{author}{Schuler, S.}, \bibinfo{author}{Adegas, F.D.},
  \bibinfo{author}{Anta, A.}, \bibinfo{year}{2017}.
\newblock \bibinfo{title}{Hybrid modelling of a wind turbine}, in:
  \bibinfo{editor}{Frehse, G.}, \bibinfo{editor}{Althoff, M.} (Eds.),
  \bibinfo{booktitle}{ARCH16. International Workshop on Applied Verification
  for Continuous and Hybrid Systems}, \bibinfo{publisher}{EasyChair}. pp.
  \bibinfo{pages}{18--26}.
\newblock \DOIprefix\doi{10.29007/tf1p}.
%Type = Article
\bibitem[{Stocco et~al.(2022)Stocco, Pulfer and Tonella}]{Stocco_2022}
\bibinfo{author}{Stocco, A.}, \bibinfo{author}{Pulfer, B.},
  \bibinfo{author}{Tonella, P.}, \bibinfo{year}{2022}.
\newblock \bibinfo{title}{Mind the gap! a study on the transferability of
  virtual vs physical-world testing of autonomous driving systems}.
\newblock \bibinfo{journal}{{IEEE} Transactions on Software Engineering} ,
  \bibinfo{pages}{1--13}\DOIprefix\doi{10.1109/tse.2022.3202311}.
%Type = Inproceedings
\bibitem[{Waga(2020)}]{Waga20}
\bibinfo{author}{Waga, M.}, \bibinfo{year}{2020}.
\newblock \bibinfo{title}{Falsification of cyber-physical systems with
  robustness-guided black-box checking}, in: \bibinfo{booktitle}{International
  Conference on Hybrid Systems: Computation and Control},
  \bibinfo{publisher}{{ACM}}. pp. \bibinfo{pages}{11:1--11:13}.
%Type = Article
\bibitem[{Wang et~al.(2022)Wang, Yu, Qiu, Sun and Farah}]{11Wang2022}
\bibinfo{author}{Wang, Y.}, \bibinfo{author}{Yu, R.}, \bibinfo{author}{Qiu,
  S.}, \bibinfo{author}{Sun, J.}, \bibinfo{author}{Farah, H.},
  \bibinfo{year}{2022}.
\newblock \bibinfo{title}{Safety performance boundary identification of highly
  automated vehicles: A surrogate model-based gradient descent searching
  approach}.
\newblock \bibinfo{journal}{IEEE Transactions on Intelligent Transportation
  Systems} , \bibinfo{pages}{1--12}\DOIprefix\doi{10.1109/TITS.2022.3191088}.
%Type = Article
\bibitem[{Wu et~al.(2022)Wu, Arcaini, Yue, Ali and Zhang}]{WuAYAZ22}
\bibinfo{author}{Wu, J.}, \bibinfo{author}{Arcaini, P.}, \bibinfo{author}{Yue,
  T.}, \bibinfo{author}{Ali, S.}, \bibinfo{author}{Zhang, H.},
  \bibinfo{year}{2022}.
\newblock \bibinfo{title}{On the preferences of quality indicators for
  multi-objective search algorithms in search-based software engineering}.
\newblock \bibinfo{journal}{Empirical Software Engineering}
  \bibinfo{volume}{27}, \bibinfo{pages}{144}.
\newblock \DOIprefix\doi{10.1007/s10664-022-10127-4}.
%Type = Inproceedings
\bibitem[{Yaghoubi and Fainekos(2019)}]{YaghoubiHSCC}
\bibinfo{author}{Yaghoubi, S.}, \bibinfo{author}{Fainekos, G.},
  \bibinfo{year}{2019}.
\newblock \bibinfo{title}{Gray-box adversarial testing for control systems with
  machine learning components}, in: \bibinfo{booktitle}{International
  Conference on Hybrid Systems: Computation and Control (HSCC)}.
%Type = Article
\bibitem[{Yamagata et~al.(2021)Yamagata, Liu, Akazaki, Duan and
  Hao}]{yamagata2020falsification}
\bibinfo{author}{Yamagata, Y.}, \bibinfo{author}{Liu, S.},
  \bibinfo{author}{Akazaki, T.}, \bibinfo{author}{Duan, Y.},
  \bibinfo{author}{Hao, J.}, \bibinfo{year}{2021}.
\newblock \bibinfo{title}{Falsification of cyber-physical systems using deep
  reinforcement learning}.
\newblock \bibinfo{journal}{IEEE Transactions on Software Engineering}
  \bibinfo{volume}{47}, \bibinfo{pages}{2823--2840}.
\newblock \DOIprefix\doi{10.1109/TSE.2020.2969178}.
%Type = Inproceedings
\bibitem[{Zeller(2017)}]{DBLP:conf/icse/Zeller17}
\bibinfo{author}{Zeller, A.}, \bibinfo{year}{2017}.
\newblock \bibinfo{title}{Search-based testing and system testing: {A} marriage
  in heaven}, in: \bibinfo{booktitle}{International Workshop on Search-Based
  Software Testing, (SBST@ICSE)}, \bibinfo{publisher}{{IEEE/ACM}}. pp.
  \bibinfo{pages}{49--50}.
%Type = Inproceedings
\bibitem[{Zhang and Arcaini(2021)}]{9Zhang2021}
\bibinfo{author}{Zhang, Z.}, \bibinfo{author}{Arcaini, P.},
  \bibinfo{year}{2021}.
\newblock \bibinfo{title}{Gaussian process-based confidence estimation for
  hybrid system falsification}, in: \bibinfo{editor}{Huisman, M.},
  \bibinfo{editor}{Pasareanu, C.S.}, \bibinfo{editor}{Zhan, N.} (Eds.),
  \bibinfo{booktitle}{Formal Methods ({FM})}, \bibinfo{publisher}{Springer}.
  pp. \bibinfo{pages}{330--348}.
%Type = Inproceedings
\bibitem[{Zhang et~al.(2021)Zhang, Lyu, Arcaini, Ma, Hasuo and
  Zhao}]{falsQBRobCAV2021}
\bibinfo{author}{Zhang, Z.}, \bibinfo{author}{Lyu, D.},
  \bibinfo{author}{Arcaini, P.}, \bibinfo{author}{Ma, L.},
  \bibinfo{author}{Hasuo, I.}, \bibinfo{author}{Zhao, J.},
  \bibinfo{year}{2021}.
\newblock \bibinfo{title}{Effective hybrid system falsification using monte
  carlo tree search guided by {QB}-robustness}, in: \bibinfo{editor}{Silva,
  A.}, \bibinfo{editor}{Leino, K.R.M.} (Eds.), \bibinfo{booktitle}{Computer
  Aided Verification}, \bibinfo{publisher}{Springer}. pp.
  \bibinfo{pages}{595--618}.
%Type = Article
\bibitem[{Zhong et~al.(2021)Zhong, Kaiser and Ray}]{4Zhong2021}
\bibinfo{author}{Zhong, Z.}, \bibinfo{author}{Kaiser, G.E.},
  \bibinfo{author}{Ray, B.}, \bibinfo{year}{2021}.
\newblock \bibinfo{title}{Neural network guided evolutionary fuzzing for
  finding traffic violations of autonomous vehicles}.
\newblock \bibinfo{journal}{CoRR} \bibinfo{volume}{abs/2109.06126}.
\newblock \URLprefix \url{https://arxiv.org/abs/2109.06126},
  \href{http://arxiv.org/abs/2109.06126}{{\tt arXiv:2109.06126}}.

\end{thebibliography}
\end{document}